\documentstyle[11pt,aaspp4]{article}

\def\etal{{\it et al.}}
\def\kms{km~s$^{-1}$}
\def\reff{r_{\rm eff}} 
\def\Reff{R_{\rm eff}} 
\def\r14{$r^{1/4}$}

\def\meanmu{\langle\mu\rangle} 
\def\meanmuK{\langle\mu_K\rangle} 
\def\meanmueff{\meanmu_{\rm eff}} 
\def\meanmueffK{\meanmuK_{\rm eff}} 
\def\Ktot{K_{\rm tot}} 
\def\Dn{D_n} 
\def\DB{D_B} 
\def\DV{D_V} 
\def\DR{D_R} 
\def\Dr{D_r} 
\def\DK{D_K} 
\def\Dsigma{$D$--$\sigma_0$}
\def\Dnsigma{$\Dn$--$\sigma_0$}

\def\DKsigma{$\DK$--$\sigma_0$}
\def\magarcsec2{mag~arcsec$^{-2}$} 
\def\rkc{$R_{\rm C}$}

\def\ikc{$I_{\rm C}$}

\def\mgtwo{Mg$_2$}

\def\mgtwosigma{\mgtwo--$\sigma_0$}

\slugcomment{submitted to The Astronomical Journal}

\lefthead{Pahre et al.}
\righthead{Near--Infrared Fundamental Plane}

\begin{document}
 
\title{Near--Infrared Imaging of Early--Type Galaxies III.  \\
	The Near--Infrared Fundamental Plane }

\author{Michael A. Pahre\altaffilmark{1,2,3,4}, S. G. Djorgovski\altaffilmark{1}, and Reinaldo R. de Carvalho\altaffilmark{1,5}}
\altaffiltext{1}{Palomar Observatory, California Institute of Technology, 
	MS 105-24, Pasadena, CA \, 91125; \emph{email:}  george@astro.caltech.edu~. } 
\altaffiltext{2}{Guest Observer, Las Campanas Observatory.} 
\altaffiltext{3}{Present address:  Harvard-Smithsonian Center for Astrophysics,
60 Garden Street, MS 20, Cambridge, MA \, 02138; \emph{email:}  mpahre@cfa.harvard.edu~. } 
\altaffiltext{4}{Hubble Fellow.} 
\altaffiltext{5}{Departamento de Astrofisica, Observatorio Nacional, CNPq, Brazil; \emph{email:}  reinaldo@maxwell.on.br~.} 

\begin{abstract}
Near--infrared imaging data on 251 early--type galaxies in clusters and groups
are used to construct the near--infrared Fundamental Plane (FP)
\begin{displaymath}
\reff \propto \sigma_0^{1.53 \pm 0.08} \langle\Sigma_K\rangle_{\rm eff}^{-0.79 \pm 0.03} .
\end{displaymath}
The slope of the FP therefore departs from the virial expectation of 
$\reff \propto \sigma_0^{2} \langle\Sigma\rangle_{\rm eff}^{-1}$ at all optical and 
near--infrared wavelengths, which could be a result of the variation of $M/L$ along 
the elliptical galaxy sequence, or a systematic breakdown of homology among the family
of elliptical galaxies.
The slope of the near--infrared FP excludes metallicity variations as the sole 
cause of the slope of the FP.
Age effects, dynamical deviations from a homology, or any combination of these 
(with or without metallicity), however, are not excluded.
The scatter of both the near--infrared and optical FP are nearly identical and
substantially larger than the observational uncertainties, demonstrating small but
significant intrinsic cosmological scatter for the FP at all wavelengths.
The lack of a correlation of the residuals of the near--infrared FP and the residuals from the
\mgtwosigma\ relation indicates that the thickness of these relations cannot be ascribed 
only to age or metallicity effects.
Due to this metallicity independence, the small scatter of the near--infrared FP excludes a
model in which age and metallicity effects ``conspire'' to keep the optical FP thin.
All of these results suggest that the possible physical origins of the FP relations
are complicated due to combined effects of variations of stellar populations and structural
parameters among elliptical galaxies.
\end{abstract}


\section{Introduction}

Correlations among the properties of elliptical galaxies have been used both as measures
of their homogeneity as a population and as indicators of the distances of individual
galaxies.
The discovery of a color--magnitude effect (\cite{baum59}) was followed by the measurement
of relative distances of galaxies and clusters using the color--magnitude relation (\cite{sandage72}; \cite{vs77};
Sandage \& Visvanathan 1978a,b). 
The relation's small scatter was used as a constraint on elliptical galaxy formation time-scales
(\cite{ble92b}).
The correlation between luminosity and velocity dispersion (\cite{fj76}) was likewise used
as a distance indicator (\cite{tonry81}; \cite{dressler84}), but Terlevich \etal\ (1981) discovered
a weak correlation between the residuals of the relation and \mgtwo, implying
that there might be a second parameter which contributes to the intrinsic scatter of the
Faber--Jackson relation.\footnote{Dressler \etal\ (1987) describe how this evidence for the
bivariate nature of elliptical galaxies found by Terlevich \etal\ (1981) might actually have
been driven more by distance errors in the Terlevich \etal\ sample, which was drawn primarily
from the field, and from a surface brightness correlation which was later found to be the
second parameter.  After correcting for surface brightness effects, Dressler \etal\ found little
or no correlation among the residuals of the Faber--Jackson relation and \mgtwo.}
Subsequent studies (\cite{dressler87}; \cite{dd87}) that included large samples of
elliptical galaxies found that surface brightness was a second parameter which caused a large 
portion of the scatter in the Faber--Jackson relation.
In this perspective, the intrinsic properties of elliptical galaxies are only found to
lie on a plane within the three--dimensional parameter space of the observables.
This Fundamental Plane (FP) is thus a set of bivariate correlations between the observed
properties of elliptical galaxies; the color--magnitude and Faber--Jackson relations are
projections of that plane onto two of the three axes.

The importance of the exact form of the slope of the FP was immediately identified
as possibly providing a strong constraint on the mass--to--light ratios ($M/L$) of 
elliptical galaxies (\cite{dressler87}; \cite{dd87}).
In particular, the virial theorem produces a prediction that 
$\reff \propto \sigma_0^2 \langle\Sigma\rangle_{\rm eff}^{-1}$ (where $\reff$ is the
effective radius, $\sigma_0$ is the central velocity dispersion, and $\langle\Sigma\rangle_{\rm eff}$
is the mean surface brightness enclosed within $\reff$) if two assumptions are made:  
(1) $M/L$ is the same for all elliptical galaxies,
and (2) elliptical galaxies form a homologous family in their scaling properties.
The latter assumption was generally taken to be true, hence virtually all researchers in
the last decade have proceeded to explore the effects of the variations of $M/L$ implied
by the FP correlations.
For example, the slope of $\Dn \propto \sigma_0^{1.2}$ from Lynden--Bell \etal\ (1988)
in the $B$--band implies that $M/L$ varies systematically with the galaxy's 
luminosity as $M/L_B \propto L_B^{0.32}$.

More recently, there have been a variety of theoretical (Capelato, de Carvalho, \& Carlberg 1995, 1997;
\cite{ciotti96}) and observational (\cite{graham97}; \cite{busarello97}) investigations 
into the question of whether or not elliptical galaxies form a homologous scaling family.
The results of these studies are not yet clear, but they seem to imply that structural
deviations of the light profiles of ellipticals from a homologous family cannot
affect the FP appreciably (\cite{graham97}), while dynamical deviations from a homologous
family can (Capelato \etal\ 1995; \cite{busarello97}; cf. \cite{graham97}).
Underlying all of these studies is an important point:  if there are significant and
systematic deviations from a homology, then these deviations should be strictly independent
of wavelength observed in constructing the global photometric parameters which enter
the FP.

In a broad sense, some of these possible implications of the FP make specific
predictions which can be tested by obtaining additional data.
For example, if the form of the FP is a direct result of a dependence of $M/L$ on
$L$ due to variations in the stellar populations parameters of age and/or metallicity,
then observations in the near--infrared should show a significantly different form
for the FP correlations as the $2.2 \mu$m light is far less sensitive than optical
light to line--blanketing and somewhat less sensitive to age effects.
Alternatively, if the origin of the FP is due to a systematic deviation of elliptical
galaxies from a homologous family, then the exact form of the FP should be independent
of wavelength.
It is also possible that some combination of these effects could be required by a
simultaneous analysis of the FP at optical and near--infrared wavelengths.

The present paper is an attempt to address these possible origins for the FP correlations
by exploring their form using near--infrared imaging data.
Early work on the FP in the $K$--band was done by Recillas--Cruz \etal\ (1990, 1991), who obtained
aperture photometry for galaxies in Virgo and Coma, and Djorgovski \& Santiago (1993),
who used aperture photometry from Persson, Frogel, \& Aaronson (1979). 
Both studies relied on optical estimates of $\reff$.
It follows directly from the existence of color gradients that $\reff$ should be smaller for
longer wavelengths and this point will be shown explicitly in a future paper (\cite{pahre98fpmodels})
for these optical and near--infrared data.
The present paper is more than just a revisiting of the $K$--band FP---it is an attempt to
study the global properties of elliptical galaxies using near--infrared photometry that is
fully independent of the optical photometry, while at the same time using a methodology for deriving 
global photometric parameters that is identical to the method at optical wavelengths.

An imaging survey of this kind and scale has only recently become possible with the introduction
of large format IR detectors ($256 \times 256$~pixel$^2$).
This project was initiated during the commissioning phase of a wide--field, near--infrared
imaging camera (\cite{murphy95}) for the Palomar 60--inch Telescope.
In the first paper of this series, early results from this survey on the $K$--band FP 
(\cite{pddc95}) indicated that there is a modest change in slope from the optical to the 
near--infrared, but not nearly as much variation as would be expected if stellar--populations 
alone were the cause of the slope of the FP (\cite{pahre97stromlo}; \cite{pahre97esoscalwork}).
The full $K$--band survey and the complete catalogs of global properties are described 
in the second paper of this series (\cite{pahre98kfpdata}) and are summarized in \S\ref{kfp-data}.
All of the data contained in the previous contributions (Pahre \etal\ 1995, 1997; 
Pahre \& Djorgovski 1997) were re--calibrated, some were re--reduced, and the global photometric 
parameters re--derived in a consistent manner as described in Pahre (1998).
The FP correlations and their many projections are derived in \S\ref{kfp-correlations}
as a way of exploring various aspects of these near--infrared data.
The \mgtwosigma\ relation is constructed for the same galaxies in \S\ref{kfp-mg2-sigma}.
Simple models will be compared to these results in \S\ref{kfp-simple}, but it will be shown
that naive models cannot explain the many observed properties of ellipticals nor are such models
unique.

\section{Description of the Data \label{kfp-data}}

The data used for this paper derive from Pahre (1998).
That paper presented near--infrared $K$--band imaging of 341 early--type galaxies,
and used those data to measure the global photometric parameters of the half--light
effective radius $\reff$, the mean surface brightness $\meanmueff$ enclosed by that
radius, the total magnitude $\Ktot$, and the diameter $\DK$ at which the mean surface
brightness, fully corrected for cosmological effects and extinction, drops to
$16.6$~mag~arcsec$^{-2}$.
The latter quantity is an analog of the $B$--band $\Dn$ parameter defined by
Dressler \etal\ (1987).
The near--infrared data were corrected for the effects of seeing.
As shown in that paper, the random uncertainties of the measured quantities are:  
0.06~dex on $\reff$; 0.21~mag on $\meanmueff$; 0.09~mag on $\Ktot$; 
0.010~dex on $\DK$; 
and 0.015~dex on $\reff - 0.32 \meanmueff$, the quantity which will enter the FP.

The galaxies in that sample are primarily drawn from nearby rich clusters of galaxies,
although additional galaxies were added from groups and the general field.
The galaxies were not selected according to any explicit criteria of completeness
(such as a magnitude--limited sample would be), but by the availability of companion
optical imaging and spectroscopic data.
The primary goal of this effort was to provide a large sample of galaxies for which
the variations of the FP correlations between the optical and near--infrared wavelengths
could be explored.
The data probe the full range of properties ($\reff$, $\Ktot$, $\sigma_0$) displayed by
the family of giant elliptical galaxies, and a significant portion of the sample is comprised
of S0 galaxies.

The optical global photometric parameters ($\reff$, $\meanmueff$, and $D_n$) and 
spectroscopic parameters (central velocity dispersion $\sigma_0$ and Magnesium line index
\mgtwo) were compiled from the literature by Pahre (1998).
All of the photometry were drawn from the $U$, $B$, $g$, $V$, $R_{\rm C}$ (or $r$), or \ikc\ 
bandpasses and converted to $V$ magnitudes for a general catalog.
Furthermore, separate catalogs were constructed for individual comparisons to preserve
the wavelength information for each optical bandpass.
The spectroscopic parameters were corrected for observed aperture size effects to a common
physical scale of $1.53 h_{75}^{-1}$~kpc; small offsets between data sets have been applied
according to prescriptions developed by other authors in the literature.
The values were then averaged to reduce the random uncertainties and minimize systematic 
errors due to some data sets.
Of the 341 galaxies imaged in the $K$--band, 95\% have velocity dispersions, 69\% have
\mgtwo\ indices, and 91\% have optical photometric parameters (either $\reff$ or $\Dn$).
The typical uncertainties for individual measurements of $\sigma_0$ and \mgtwo\ are
$0.04$~dex and $0.013$~mag (\cite{smith97}), respectively.
A substantial fraction of the entire sample has more than one measurement of these parameters which
were then averaged, so these two uncertainty estimates can be taken as a universal upper limit
to the measurement uncertainties.

Many, if not most, of the literature sources suggest that their velocity dispersions are less
reliable below $100$~km~s$^{-1}$, but a bias in the slope of the FP can be introduced by 
imposing a cut on $\sigma_0$.
It will be important to investigate what effect changing the lower cutoff for $\sigma_0$
has on the slope and scatter of the FP.
Small measures of the effective radius, such as $\reff \leq 1$~arcsec (the median seeing was
$1.34$~arcsec FWHM), have large random uncertainties and probably substantial systematic 
errors, and should probably also be discarded.
In the sample, 4\% of the galaxies have morphological type SB0 or later, and another 4\% have
S0/a type; caution should be exercised when studying the global properties of these galaxies.
One galaxy (D45 in cluster Abell~194) appears to be a misidentification either in
the optical or near--infrared as evidenced by its color $(V-K)=1.28$~mag, which appears to
be much too blue compared to the mean $(V-K)=3.15$~mag for the entire sample.
Five galaxies in the Virgo cluster were removed from the sample, as their accurate distances 
as derived by the surface brightness fluctuations method (provided by J. Blakeslee and J. Tonry;
see \cite{tonry97}) show that they are either in the background W Cloud (NGC~4168, NGC~4261,
and NGC~4365) or the foreground (NGC~4660 and NGC~4697).

\section{Analysis of the Elliptical Galaxy Correlations \label{kfp-correlations}}

\subsection{The Near--Infrared Fundamental Plane \label{kfp-kfp}}

Galaxies were drawn from the sample described in \S\ref{kfp-data}.
Only those galaxies residing in a cluster or group with four or more observed galaxies were
included in the FP fits, resulting in 16 clusters/groups and 249 galaxies with $\sigma_0 > 1.8$.
Two of the five Leo~I Group galaxies were excluded as a result of these selection criteria,
although the remaining three galaxies were retained in the sample for completeness.

The ``standard'' FP equation is usually written as
\begin{equation}
	\log \reff ({\rm arcsec}) = a \log \sigma_0 ({\rm km~s}^{-1}) + b \meanmueff ({\rm mag~arcsec}^{-2}) + c_i
	\label{kfp-eq-fpgen}
\end{equation}
where $a$ is usually identified as the ``slope'' of the FP and $c_i$ are the ``intercepts.''
The intercept of the relation will vary with distance.
The galaxies in each of the 16 groups and clusters are assumed to lie at the same distance, hence
there are $i=16$ different intercepts $c_i$.

The Equation \ref{kfp-eq-fpgen} was fit by minimizing the sum of the absolute deviations of the points orthogonally
from the relation using the program GAUSSFIT (\cite{gaussfit}).
During the first iteration, two points which are outliers (D9 in Cen30 and S201 in Hydra)
were identified and excluded from the analysis.
The resulting FP in the near--infrared $K$--band is
\begin{equation}
	\begin{array}{rcllll}
    \log \reff ({\rm arcsec}) & = &\, \, 1.53 \log \sigma_0 + &\, \, 0.314 \meanmueffK + c_i & N=251 & {\rm rms} = 0.096 {\rm~dex} \\
                              &   &  \pm 0.08                 &  \pm 0.011
	\end{array}
	\label{kfp-eq-kfp-all}
\end{equation}
The uncertainties in the coefficients were determined by 100 iterations of bootstrap 
resampling of the data-points.
The individual intercepts $c_i$ for the fit, and the rms about the fit for each cluster or group, are listed
in Table~\ref{tab1}.
Since the rms is quoted in units of $\log \reff$, the uncertainty on each intercept is therefore
${\rm rms}/\sqrt{N-1}$.
The relation in Equation~\ref{kfp-eq-kfp-all} is equivalent to the scaling relation
$\reff \propto \sigma_0^{1.53 \pm 0.08} \langle\Sigma_K\rangle_{\rm eff}^{-0.79 \pm 0.03}$.

\placetable{tab1}


Changing the lower cutoff for $\sigma_0$ from $1.8$~dex to $2.0$~dex changes the value 
of $a$ by $\leq 0.01$~dex, changes $b$ by $\leq 0.001$~dex, reduces the scatter 
by 10\% to $0.089$~dex, and excludes 23 galaxies (9\% of the total).
Hence, the solution to the FP is robust to the lower $\sigma_0$ cutoff, although the galaxies with
the lowest $\sigma_0$ appear to contribute the largest to the observed scatter.
Minimizing the orthogonal variance (instead of the absolute value of the
deviation) from the FP relation with $\sigma_0 \geq 1.8$~dex results in a small change in the 
slope of the FP to:
\begin{equation}
	\begin{array}{rcllll}
    \log \reff ({\rm arcsec}) & = &\, \, 1.63 \log \sigma_0 + &\, \, 0.320 \meanmueffK + c_i & N=251 & {\rm rms} = 0.099 {\rm~dex} \\
                              &   &  \pm 0.06                 &  \pm 0.008
	\end{array}
	\label{kfp-eq-kfp-variance}
\end{equation}
Since the coefficients in Equations~\ref{kfp-eq-kfp-all} and \ref{kfp-eq-kfp-variance} are 
equivalent within the uncertainties, the fit to the FP is insensitive to the exact fitting method.
The method of minimizing the absolute value of the orthogonal deviation from the fit is to
be preferred, however, as it is less sensitive to outliers.

\begin{figure}
	\epsscale{1.0}
	\plotone{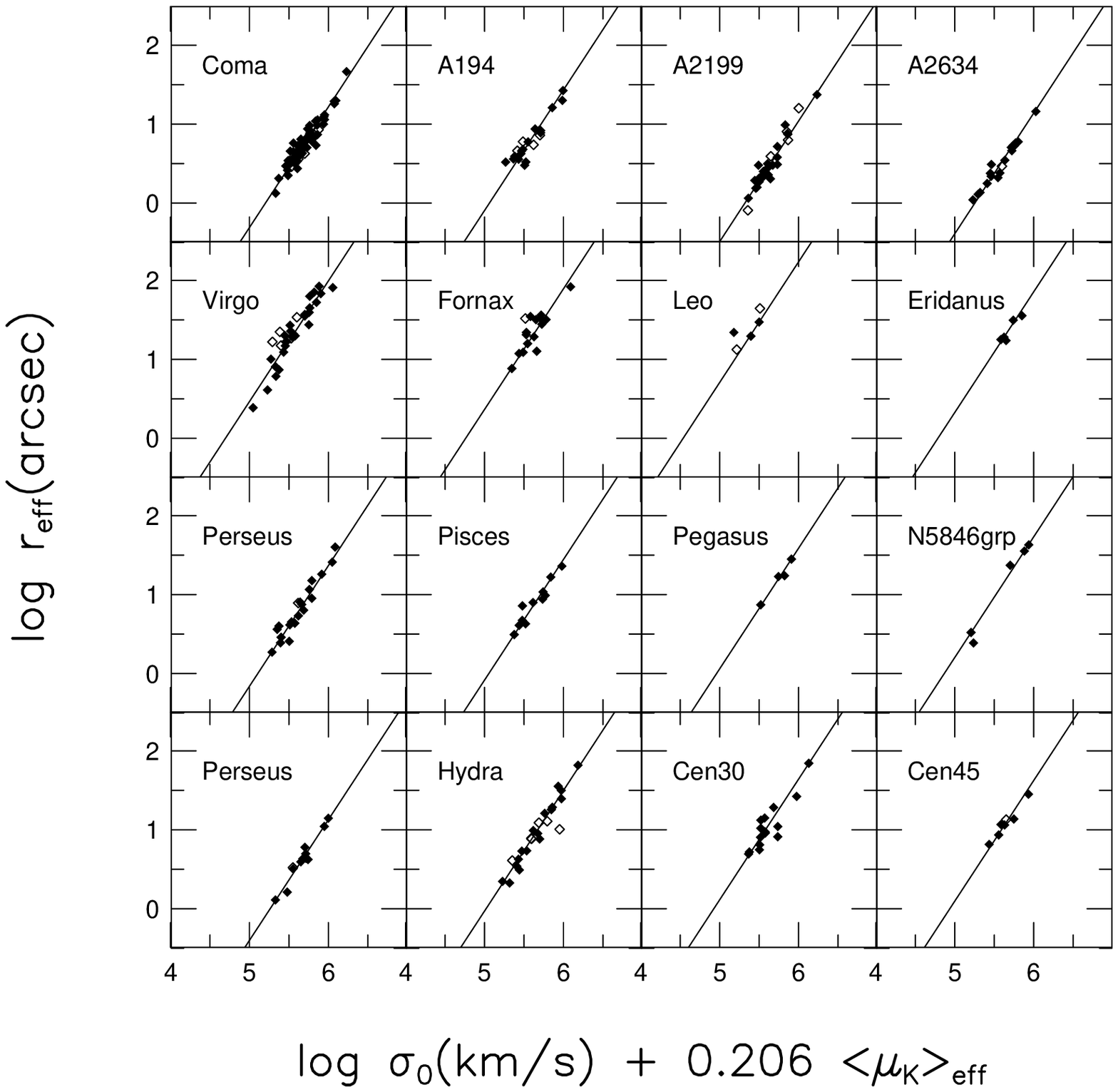}
	\caption[The Fundamental Plane (FP) in the $K$--band for Each Cluster]{
	The Fundamental Plane in the near--infrared for the 16 clusters and groups 
	in the simultaneous fit represented by the solution of Equation~\ref{kfp-eq-kfp-all} and
	the intercepts in column 2 of Table~\ref{tab1}.
	The FP is described by the scaling relation
	$\reff \propto \sigma_0^{1.53} \langle\Sigma_K\rangle_{\rm eff}^{-0.79}$
	with a scatter of $0.096$~dex in $\log \reff$; the scatter is reduced by 10\%
	of the galaxies with $\sigma_0 < 100$~km~s$^{-1}$ are excluded.
	The fitted galaxies are plotted as solid symbols, while those excluded from
	the fit ($\log \sigma_0 < 1.8$, late--type morphology, $\reff < 1$~arcsec, or 
	background/foreground in the Virgo cluster) are plotted as open symbols.
	The FP fit is plotted in each panel as a solid line.
	\label{fig-kfp-panels}
	}
\end{figure}

The simultaneous fit to all clusters is displayed in Figure~\ref{fig-kfp-panels} with the 
data subdivided into the 16 individual clusters or groups.
It is clear from this figure that the simultaneous fit is a representative description of the
properties of the early--type galaxies in all of the clusters.
There is no clear deviation from this mean relation.

\placefigure{fig-kfp-panels}

The 11 clusters with more than ten galaxies were fit individually to Equation~\ref{kfp-eq-fpgen}
as a test of the universality of the FP relation.
The difficulty with all 11 separate fits, however, is that the number of galaxies in each cluster is
small enough that the slope of the relation is not accurately determined.
Once again, the uncertainties on the fitted coefficients $a$ and $b$ have been determined
using the bootstrap procedure.
In eight of the 11 cases, the fits have $a$ within one standard deviation of the value
$a=1.53$ from the simultaneous fit, suggesting both that the uncertainties are reasonably
estimated and that better fits are limited by the number of galaxies per cluster.
These individual fits are listed in Table~\ref{tab1}.

The same 11 clusters were also fit individually by constraining $b=0.314$ from the previous 
simultaneous fit.
This is possible because virtually every study of the FP (optical and near--infrared)
obtains a similar value for this parameter, hence it should be possible to constrain its
value {\sl a priori}.
These fits show significantly smaller uncertainty in their determination of $a$ than the
unconstrained fits, and are listed in Table~\ref{tab1}.
In this case, seven of the 11 clusters have a slope $a$ within one standard deviation of the
value from the simultaneous fit.

\begin{figure}
	\epsscale{0.7}
	\plotone{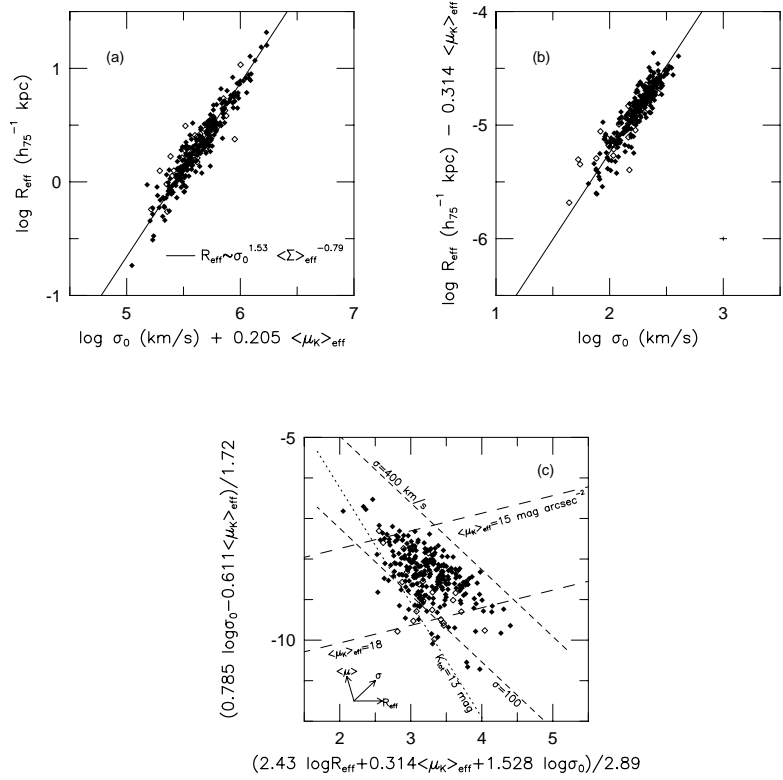}
	\caption[The FP in the $K$--band in Edge--On and Face--On Per\-spec\-tives]{
	(a)  The Fundamental Plane in the $K$--band for the combined 16 cluster and group sample,
	seen edge--on along its long side.
	The symbols are as in Figure~\ref{fig-kfp-panels}.
	The ordinate is in units of kpc assuming $H_0 = 75$~km~s~Mpc$^{-1}$.
	(b)  The FP in the $K$--band for the combined 16 cluster and group sample,
	seen edge--on along its short side.
	In this view of the FP, the observationally--correlated measurement errors in $\Reff$
	and $\meanmueffK$ (ordinate) are separated from the independent measurement uncertainties in
	$\sigma_0$ (abscissa); the typical measurement uncertainties 
	are shown in the lower right--hand corner of the panel.
	(c)  The FP seen face--on.  Galaxies do not uniformly populate this planar surface.
	While the $K$--band data in this paper are not drawn from a strictly magnitude--limited
	sample, they do behave as though a $\Ktot \lesssim 13$~mag limit (dotted line) were imposed.
	Most of the galaxies are found to have $15 < \meanmueffK < 18$~\magarcsec2\ (long--dashed lines),
	although there are no clear selection effects causing this distribution of galaxy properties.
	Furthermore, there are no galaxies with properties in the upper--right portion of the figure,
	which could be caused by the lack of galaxies with velocity dispersions $\sigma_0 > 400$~\kms\
	(short--dashed line).
	The vectors drawn in the lower left--hand corner show the direction in which each of the
	observed quantities varies along the plane.
	\label{fig-kfp-combined}
	}
\end{figure}

The adopted form of the FP in Equation~\ref{kfp-eq-kfp-all} is plotted for all 301 galaxies in
these 16 clusters and groups in Figure~\ref{fig-kfp-combined}, both in face--on and edge--on
perspectives.
While the edge--on view with $\log\Reff$ as the ordinate\footnote{In this paper, a distinction will
be made between the angular effective radius $\reff$, measured in arcsec, and the effective radius
$\Reff$ in physical scale, measured in kpc.} 
is the most common method of displaying the FP, the edge--on view with $\log\Reff - 0.314 \meanmueff$ 
as the ordinate is easier to interpret.
Virtually every study of the FP (optical and near--infrared) obtains the same 
relationship between $\Reff$ and $\meanmueff$, but there may be significant variation
in the relationship between $(\log\Reff - 0.314 \meanmueff )$ and $\log\sigma_0$, depending
on wavelength.
Furthermore, this edge--on perspective of the FP, seen from its short side, separates
the correlated measurement errors in $\reff$ and $\meanmueff$ from the independent measurement
errors in $\sigma_0$.
The FP in physical units as plotted in Figure~\ref{fig-kfp-combined} is 
$\log\Reff ( h_{75}^{-1} {\rm~kpc} ) = 1.53 \log\sigma_0 + 0.314 \meanmueff - 8.30$.

\placefigure{fig-kfp-combined}

In the face--on view of the FP in Figure~\ref{fig-kfp-combined}(c), it is seen that 
galaxies do not uniformly populate this planar surface.
While the $K$--band data in this paper are not drawn from a strictly magnitude--limited
sample, they do behave as though a $\Ktot \lesssim 13$~mag limit were imposed.
Most of the galaxies are found to have $15 < \meanmueffK < 18$~\magarcsec2\ (long--dashed lines),
although there are no clear selection effects causing this distribution of galaxy properties.
Furthermore, there are no galaxies with properties in the upper--right portion of the figure,
which could be caused by the lack of galaxies with central velocity dispersions $\sigma_0 > 400$~\kms, 
although there is no selection limit imposed on this portion of the FP.

\subsection{The \DKsigma\ Relation \label{kfp-dk-sigma}}

Dressler \etal\ (1987) introduced a parameter $D_n$ which was defined as the diameter at which
the circular mean surface brightness (fully corrected for cosmological effects and Galactic
extinction) dropped to a fiducial value.
This parameter was chosen, in effect, to be a combination of the $\reff$ and $\meanmueff$ terms
in the FP correlations, thereby simplifying the FP to a \Dnsigma\ relation.
They defined this fiducial surface brightness to be $20.75$~\magarcsec2\ in the $B$--band:
the surface brightness was faint enough that $\DB$ was typically much larger than the seeing
disk, while bright enough that interpolation (rather than extrapolation)
was used to evaluate $\DB$ from their aperture magnitude data.\footnote{The notation of $\DB$ 
will be adopted for the rest of the paper to distinguish the $\Dn$ parameter as defined in the $B$--band from the
equivalent diameter as defined in another bandpass.  The name ``\Dsigma\ relation'' will refer
to the correlation in all bandpasses.}

Lucey \& Carter (1988) defined an equivalent $\DV$ parameter in the $V$--band to be the 
diameter at which the mean surface brightness drops to 19.8~\magarcsec2\ (this assumes a mean
galaxy color of $(B-V)_0=0.95$~mag), Smith \etal\ (1997) defined $\DR$ for the \rkc--band to be
at $\meanmueff = 19.23$~\magarcsec2\ (assuming $(V-R_{\rm C})=0.57$~mag), J\o rgensen \etal\
(1995a) defined the Gunn $r$--band $\Dr \equiv 2r_n$ to be at $\meanmueff = 19.6$~\magarcsec2\ 
(assuming $(V-r)=0.2$~mag), and Pahre (1998) defined the $K$--band $\DK$ 
to be at $\meanmueff = 16.6$~\magarcsec2\ (assuming $(V-K) = 3.2$~mag).
By using typical colors for early--type galaxies in constructing these definitions, the average
value of $D$ measured for a sample of galaxies should be approximately independent of bandpass.
The slope of the \Dsigma\ relation may vary between bandpasses, however, causing there to
be a systematic variation of the $D$ parameter {\sl as a function of $\sigma$} in different 
bandpasses (while keeping the mean $D$ similar).
For example, if $\DV \propto \sigma^{a_V}$ and $\DK \propto \sigma^{a_K}$,
then $\log \DV - \log \DK = (a_V - a_K) \log\sigma + {\rm~constant}$.

\begin{figure}
	\epsscale{1.0}
	\plotone{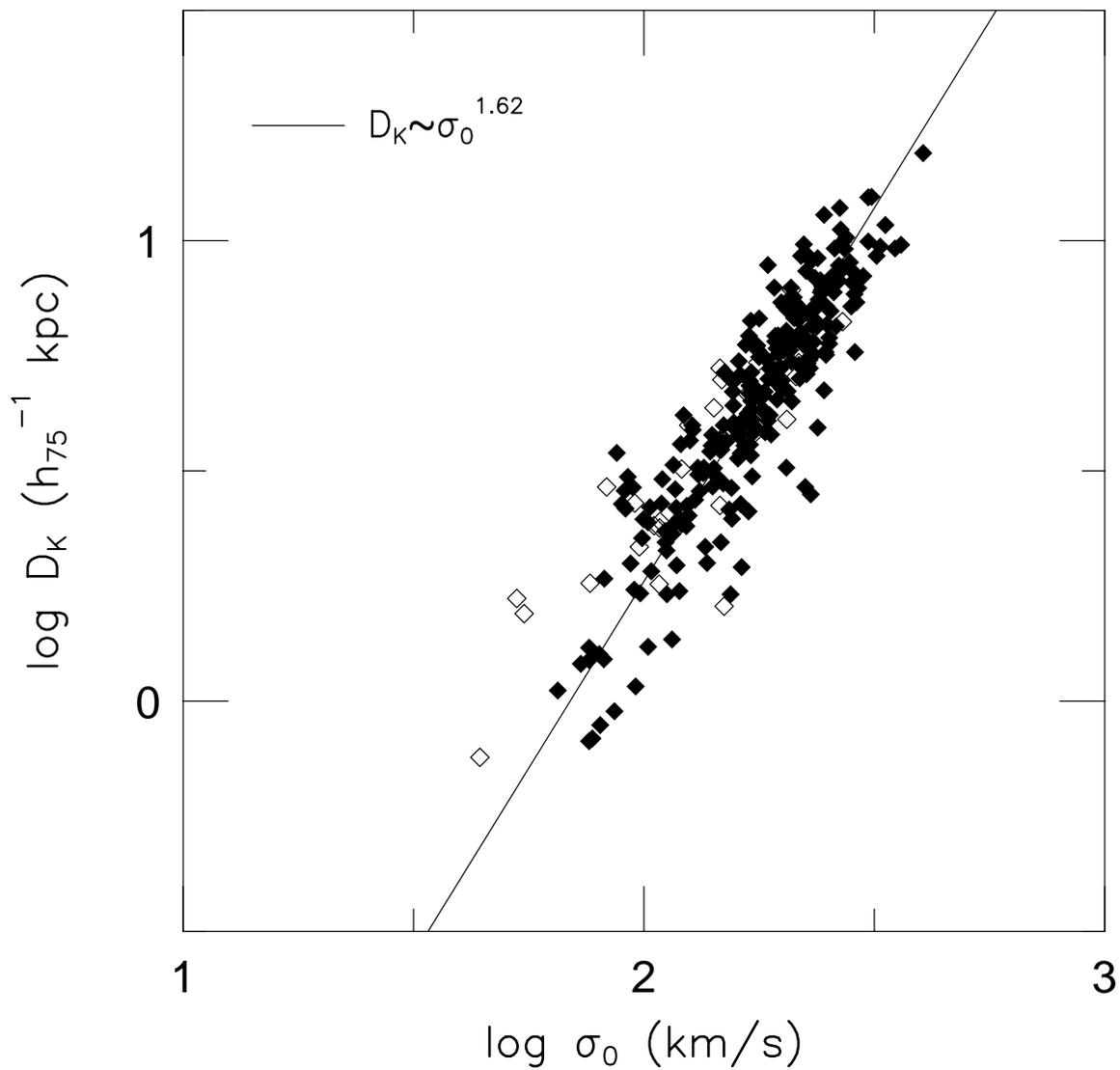}
	\caption[The \DKsigma\ Relation]{
	The \DKsigma\ relation for the galaxies in the 16 clusters and groups of the survey, plotted
	in physical units.
	The scatter for this relation is 0.112~dex in $\log\DK$, which is slightly higher than the FP
	itself, since the \DKsigma\ is not quite an edge--on view of the FP.
	\label{fig-kfp-dksigma}
	}
\end{figure}

The \DKsigma\ relation was fit for the galaxies in the same 16 clusters and groups as in 
\S\ref{kfp-kfp}, excluding galaxies using similar criteria 
($D_K < 2$~arcsec, $\log \sigma_0 < 1.8$, late types), and using the bootstrap method to
estimate uncertainties in the fitted coefficients.
The best fitting relation is
\begin{equation}
	\begin{array}{rclll}
	\log \DK ( h_{75}^{-1} {\rm~kpc} )	& = &\, \, 1.62 \log \sigma_0 - 2.984 & N = 252 & {\rm rms} = 0.112 {\rm~dex}  \\
						&   &  \pm 0.07
	\end{array}
	\label{kfp-eq-dksigma}
\end{equation}
and is displayed in Figure~\ref{fig-kfp-dksigma}.
The slope of this relation is consistent, given the estimated uncertainties, with the
full FP relation in Equations~\ref{kfp-eq-kfp-all} and \ref{kfp-eq-kfp-variance}.
The scatter of the \DKsigma\ relation, however, is 15\% higher than the $K$--band FP, despite the
fact that the measurement uncertainty of $\DK$ is actually smaller than that for
$\reff - 0.32\meanmueff$ (\cite{pahre98kfpdata}).
This is most likely due to the fact that the \DKsigma\ relation is nearly, but not quite, viewing the
FP edge--on.

\placefigure{fig-kfp-dksigma}

\subsection{The FP As Seen in $\kappa$--Space \label{kfp-kappa} }

\begin{figure}
	\epsscale{0.7}
	\plotone{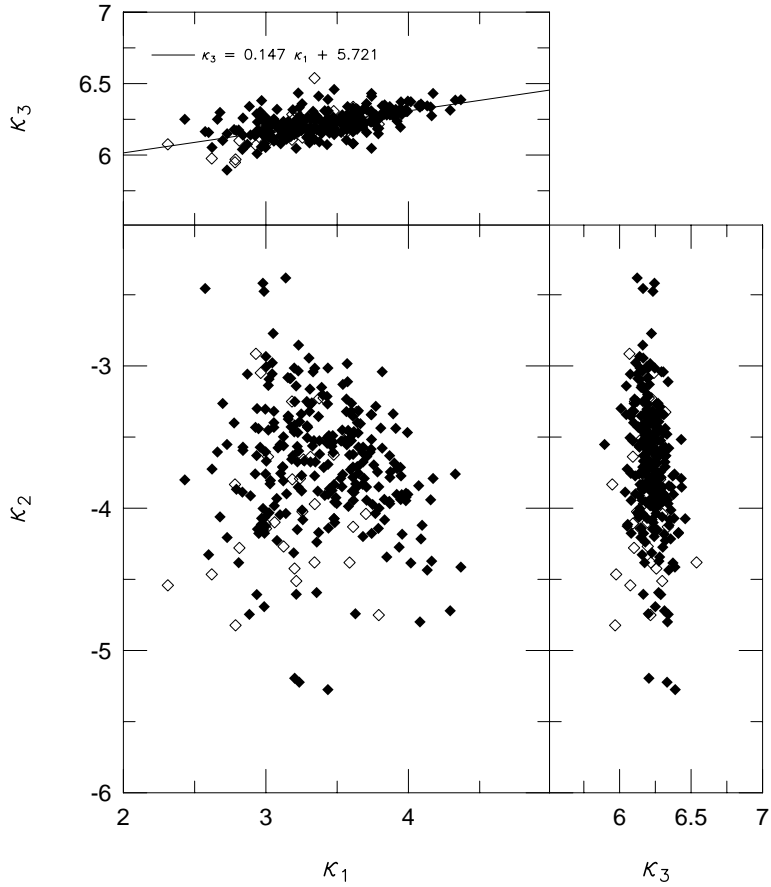}
	\caption[The $K$--band FP Viewed in $\kappa$--Space]{
	The $K$--band FP viewed in the $\kappa$--space perspective.
	This coordinate system (defined by \cite{bender92}) is given in Equation~\ref{kfp-eq-kappa-def},
	and was designed such that $\kappa_1$ is roughly proportional to the logarithm of mass and
	$\kappa_3$ is roughly proportional to the logarithm of mass--to--light ratio.
	The fit in the top panel corresponds to the ``observed'' scaling relations 
	$(M/L_K) \propto M^{0.15 \pm 0.01}$ or $(M/L_K) \propto L_K^{0.17 \pm 0.01}$,
	under the assumptions that there are no color gradients in elliptical galaxies and dynamical
	homology is preserved within the family of elliptical galaxies.
	\label{fig-kfp-kappa}
	}
\end{figure}

Since elliptical galaxies only populate a plane in the three--dimensional space of the observables
$(\Reff,\meanmueff,\sigma_0)$, it is straightforward to construct an orthogonal transformation
from this observed coordinate system to another one which might
facilitate a physical interpretation of the underlying parameters.
Since there are many possible transformations to accomplish this, the chief practical difficulty
is identifying a transformation which permits a robust and unbiased physical interpretation
of the true galaxy properties (\cite{sgd88}).
A transformation of this kind was proposed by Bender, Burstein, \& Faber (1992):
\begin{equation}
	\begin{array}{rcl}
    	\kappa_1 	& \equiv 	& ( 2 \log \sigma_0 + \log \Reff ) / \sqrt{2} \\
    	\kappa_2 	& \equiv 	& ( 2 \log \sigma_0 + 0.8 \meanmueff - \log \Reff ) / \sqrt{6} \\
    	\kappa_3 	& \equiv 	& ( 2 \log \sigma_0 + 0.4 \meanmueff - \log \Reff ) / \sqrt{3}
	\end{array}
	\label{kfp-eq-kappa-def}
\end{equation}
where the quantities $\kappa_i$ were constructed with the hope that $\kappa_1$ would be proportional to mass,
$\kappa_3$ would be proportional to mass--to--light ratio, and $\kappa_2$ (which is required to be orthogonal
to $\kappa_1$ and $\kappa_3$) would be proportional to the product of mass--to--light ratio and
the third power of mean surface brightness.
This ``$\kappa$--space'' is displayed in Figure~\ref{fig-kfp-kappa} for the $K$--band survey.
Given the above interpretation of $\kappa_1$ and $\kappa_3$, the fitted line between these
two variables
\begin{equation}
	\begin{array}{rclllll}
	\kappa_3 	& = &\, \, 0.147 \kappa_1 & + &\, \, 5.721 \meanmueffK & N=251 & {\rm rms} = 0.068 {\rm~dex} \\
                              &   &  \pm 0.011                 & &  \pm 0.038
	\end{array}
	\label{kfp-eq-kband-kappa}
\end{equation}
then implies that the ``observed mass--to--light ratio''\footnote{For the reasons discussed later in this
section, we prefer to distinguish between this observed relationship (between $\kappa_1$ and $\kappa_3$)
and the {\sl intrinsic} $M$ and $(M/L)$; the intrinsic properties may or may not be fully described by 
the axes $\kappa_1$ and $\kappa_3$.} 
in the $K$--band varies as $(M/L_K) \propto M^{0.147 \pm 0.011} \propto L_K^{0.172 \pm 0.013}$.
This conclusion is dependent on elliptical galaxies forming a dynamically homologous
family in which the central velocity dispersion scales to the effective velocity
dispersion (the velocity dispersion within the effective radius) independently
of the mass or luminosity of the galaxy.
The uncertainties in Equation~\ref{kfp-eq-kband-kappa} were derived using bootstrap 
resampling of the data points.
The cumulative observational uncertainties in this equation are $0.033$~dex in $\kappa_3$,
which is substantially smaller than the rms of the fit, implying a substantial intrinsic
scatter of the ``observed mass--to--light ratio'' for any given ``mass.''

\placefigure{fig-kfp-kappa}

While it is desirable to choose an orthogonal coordinate system which might directly relate
the observables to underlying physical properties of elliptical galaxies, this conceptualization
of the FP has a number of problems.
First, the quantity $\Reff$ is not equivalent to $R_m$, the half--mass radius, but instead
varies with the observed wavelength.
This generally follows from the presence of color gradients in elliptical galaxies 
(e.g., \cite{franx89}; Peletier \etal\ 1990a,b), but will be shown explicitly for the case
of comparing $V$--band and $K$--band effective radii in the next paper in this series
(\cite{pahre98fpmodels}).
It follows that the value of $\kappa_1$, which was intended to create a quantity which is proportional
to mass, {\sl systematically varies} as a function of wavelength while mass, of course, does not.
Using the terminology of Djorgovski \etal\ (1988), the mapping from $\Reff$ to gravitational radius $R_g$ 
is accomplished in the equation $\Reff = k_R R_g$; the argument here is that the structure function
$k_R$ is not a constant, as is implicitly assumed in the construction of $\kappa$--space by 
Bender \etal\ (1992), but is instead a function of wavelength.

Second, while the evidence is not yet strong, the lowest luminosity ellipticals show no detectable 
color gradients (\cite{peletier90a}), suggesting that there could be a dependence of the size of the
color gradient on luminosity (and hence mass).
The size of the color gradients also depends on the wavelength sampled (Franx \etal\ 1989; 
Peletier \etal\ 1990a,b).
The mappings from $\kappa_1$ to mass and $\kappa_3$ to mass--to--light ratio are therefore a 
function of both wavelength and size of the color gradient (which is, in turn, a function of mass).
Again using the terminology of Djorgovski \etal\ (1988), this means that the composite structure
functions $k_{SR}$ and $k_{SL}$ (see their Equations 5--8) are also not constant, as is implied by the
construction of $\kappa$--space, but are instead functions of wavelength and mass.

Third, the use of the central velocity dispersion $\sigma_0$ in deriving mass at the effective radius
assumes dynamical homology among elliptical galaxies when mapping $\sigma_0$ (the central velocity
dispersion) to $\sigma_{\rm eff}$ (the velocity dispersion within the effective radius).
Whether or not the internal stellar velocity distributions of elliptical galaxies form a homologous family 
is a point of considerable debate. 
Empirical arguments (\cite{jfk95b}, \cite{busarello97}) 
and numerical simulations of dissipation-less merging (Capelato \etal\ 1995) seem to suggest 
that the way $\sigma_0$ scales to $\sigma_{\rm eff}$ is a function of galaxy mass or luminosity.
The mapping from $\kappa_1$ to mass and $\kappa_3$ to mass--to--light ratio are therefore a function
of mass or luminosity, and possibly a function of other physical processes which are currently
poorly understood.
Once again using the terminology of Djorgovski \etal\ (1988), the mapping from the intrinsic 
velocity distribution $\langle v^2 \rangle$ to the observed $\sigma$ in the equation 
$\sigma^2 = k_V \langle v^2 \rangle$ is accompanied by a dependence of $k_V$ on galaxy mass (or
luminosity) and systematic deviations from a dynamical homology.

In summary, because the mapping from the observables to $\kappa$--space is a {\sl function} 
of wavelength, luminosity, and deviations from dynamical homology, 
and furthermore because the mapping from $\kappa$--space to mass and mass--to--light ratio is also 
a {\sl function} of wavelength, luminosity, and deviations from dynamical homology,
we eschew the use of $\kappa$--space since it is an obfuscation, rather than an illumination,
of the fundamental physical quantities of elliptical galaxies which we wish to understand.
At its best, the $\kappa$--space formalism is merely an \emph{intermediate} orthogonal transformation
between the observables $\left[\Reff,\meanmueff,\sigma_0\right]$ and the desired physical properties
$\left[M,L,M/L\left(\lambda\right)\right]$.

\subsection{The \mgtwosigma\ Relation \label{kfp-mg2-sigma}}

The \mgtwosigma\ relation is a correlation between two distance independent quantities and
hence useful both as a diagnostic and as a constraint on formation processes for elliptical
galaxies as a family.
Guzm\'an (1995), for example, found that the residuals of the \mgtwosigma\ relation
and the \Dsigma\ relation showed systematic differences between the Hydra--Centaurus region
and the Coma cluster, thereby suggesting that there are differences between the global
properties of galaxies in those two environments.

\begin{figure}
	\epsscale{1.0}
	\plotone{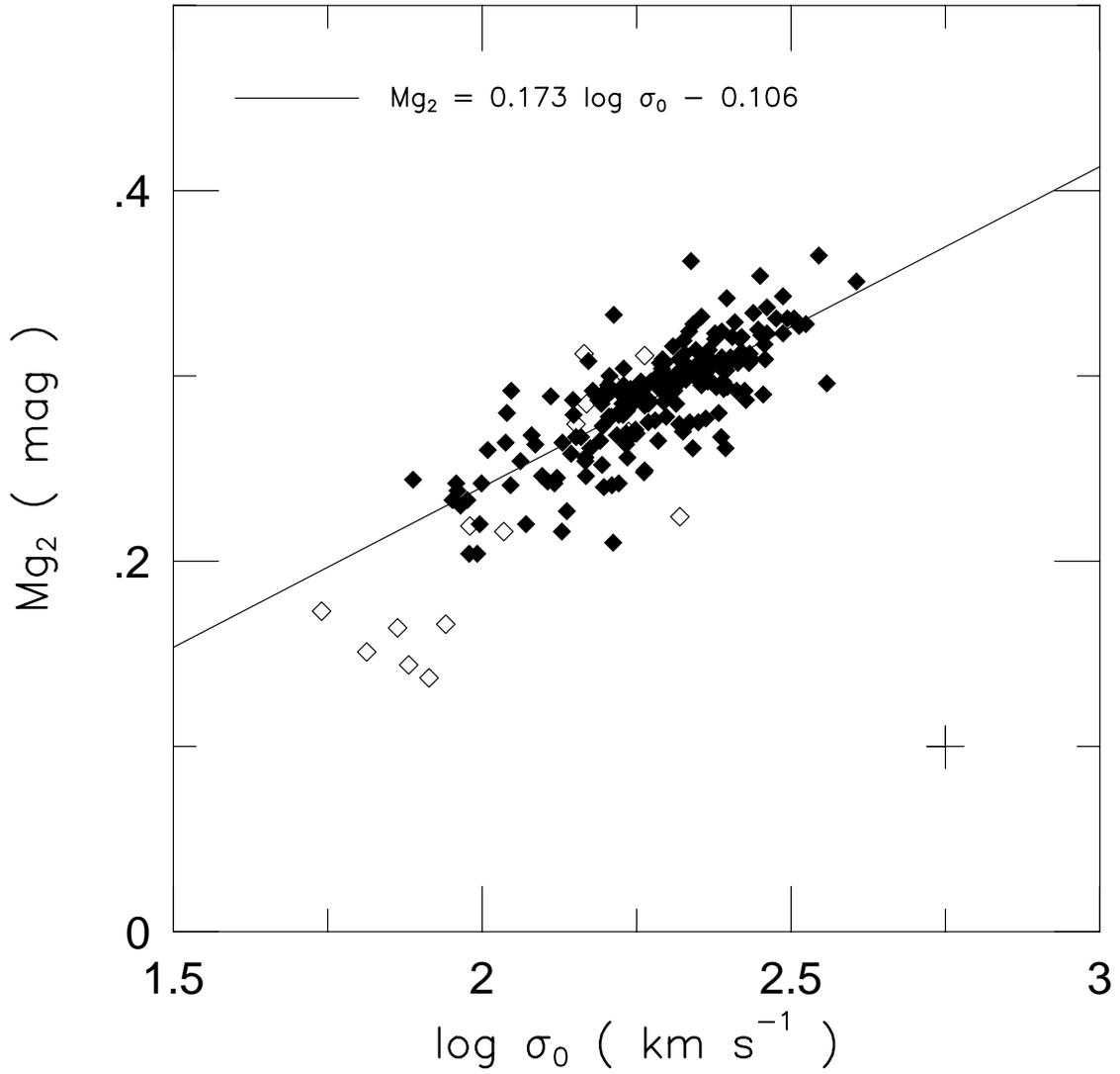}
	\caption[The \mgtwosigma\ Relation]{
	The \mgtwosigma\ relation for the 182 galaxies with \mgtwo\ measurements.
	The scatter about this relation is $0.019$~mag in \mgtwo, which is significantly
	larger than the typical measurement uncertainties of $0.013$~mag (shown in lower
	right--hand corner of figure).
	\label{fig-kfp-mg2-sigma}
	}
\end{figure}

Only a fraction of galaxies for which $\sigma_0$ is available also have \mgtwo\ values available.
Of the entire sample of 301 early--type galaxies in these 16 clusters and groups, only 188 galaxies
(62\%) fit the criteria of $\log\sigma_0 \geq 1.8$ and have \mgtwo\ measurements.
There are six galaxies at low $\sigma_0$ that show anomalously low \mgtwo\ and are therefore
excluded:  M32, NGC~3489 in the Leo Group, and NGC~4239, NGC~4468, NGC~4476, 
and NGC~4733 in the Virgo cluster (NGC~4489 was previously excluded for $\log \sigma_0 < 1.8$).
Many of these are dwarf galaxies which are known to follow different FP correlations.
The criterion used for this exclusion was that all galaxies satisfy
Mg$_2 < {5 \over 7} ( 2.2 - \log\sigma_0 )$.
The best fitting \mgtwosigma\ relation is
\begin{equation}
	\begin{array}{rcllll}
	{\rm Mg}_2 ( {\rm~mag} )	& = &\, \, 0.173 \log \sigma_0 	& - 0.106 & N = 182 & {\rm rms} = 0.019 {\rm~mag}  \\
					&   &  \pm 0.010		& \pm 0.024
	\end{array}
	\label{kfp-eq-mg2sigma}
\end{equation}
This relation is plotted in Figure~\ref{fig-kfp-mg2-sigma}.
The slope of this relation is slightly shallower than the value of $0.196 \pm 0.009$ found by 
J\o rgensen (1998), and the scatter is slightly smaller than the $0.025$~mag of J\o rgensen.
Inclusion of the six galaxies anomalously low in \mgtwo\ and the one galaxy with $\log \sigma < 1.8$
produces a \mgtwosigma\ relation Mg$_2 = 0.188 \pm 0.012 \log\sigma_0 - 0.140 \pm 0.026$ 
with a scatter of $0.021$~mag; this is closer to, and statistically indistinguishable from,
J\o rgensen's results.
The form in Equation~\ref{kfp-eq-mg2sigma} will be used, however, as it best represents the
properties of the normal elliptical galaxies.

\placefigure{fig-kfp-mg2-sigma}

\subsection{The \mgtwo\ Near--Infrared Fundamental Plane \label{kfp-mg2-kfp}}

An alternative form of the FP was proposed by de Carvalho \& Djorgovski (1989):  substitute
a stellar populations indicator, such as the \mgtwo\ index, for the dynamical or mass indicator
$\sigma_0$ in the FP relation.
The motivation for this is that since \mgtwo\ and $\sigma_0$ are strongly correlated with each other
(as shown above in \S\ref{kfp-mg2-sigma}) then metallicity could actually be the fundamental
physical property that causes the slope of the FP to deviate from the virial expectation.
de Carvalho \& Djorgovski also showed that a metallicity sensitive color could be substituted for $\sigma_0$, although
that approach will not be pursued here due to the heterogeneity of the derived optical--infrared
colors in Paper II (see \cite{pahre98kfpdata}).\footnote{Basically, the fundamental requirement for such an 
investigation of the ``color FP'' is to understand the difference between an ``aperture'' color--magnitude relation, 
the standard form which relies on a color measured in a fixed physical aperture size for all galaxies, and a ``global'' 
color--magnitude relation, which is evaluated at some fiducial scaling radius.  Part of the 
slope of the ``aperture'' color--magnitude relation is certainly due to the presence of color
gradients which act in the sense that ellipticals are redder in their centers:  the smallest galaxies
have their colors evaluated at large $r/\reff$ where their color is bluer, while the largest galaxies
have their colors evaluated at small $r/\reff$ where their color is redder.  Future work should
explicitly distinguish between the two effects of color gradients and global color differences in
order to place a constraint on the global properties of ellipticals.}
The resulting FP in the near--infrared $K$--band using the \mgtwo\ index in place of $\log\sigma_0$ is
\begin{equation}
	\begin{array}{rcllll}
	\log \reff ({\rm arcsec}) & = &\, \, 8.3 {\rm~Mg}_2 + &\, \, 0.324 \meanmueffK + c_i & N=181 & {\rm rms} = 0.172 {\rm~dex} \\
	                          &   &  \pm 0.9              &  \pm 0.015
	\end{array}
	\label{kfp-eq-mg2-kfp}
\end{equation}
The slope of this relation is $8.3 \pm 0.9$, as predicted by the slope of the \mgtwosigma\ relation combined
with the standard form of the near--infrared FP, i.e., $(1.53 \pm 0.08) / (0.173 \pm 0.010) = 8.8 \pm 0.7$.
The observational uncertainties increase when \mgtwo\ is used instead of $\log\sigma_0$ since the
latter quantity has $\sim 3$ times greater uncertainty, while the slope has changed by more than
a factor of five from Equation~\ref{kfp-eq-kfp-all} to Equation~\ref{kfp-eq-mg2-kfp}.
The scatter in the near--infrared \mgtwo\ FP, however, has increased by much more than this difference,
suggesting that \mgtwo\ is not nearly as good an indicator as the velocity dispersion in describing the
fundamental and regular physical properties in elliptical galaxies that give rise to the FP.
The \mgtwo\ index could be identifying real differences in the stellar populations among 
galaxies and hence shows larger scatter when it is substituted into the FP.

\begin{figure}
	\epsscale{1.0}
	\plotone{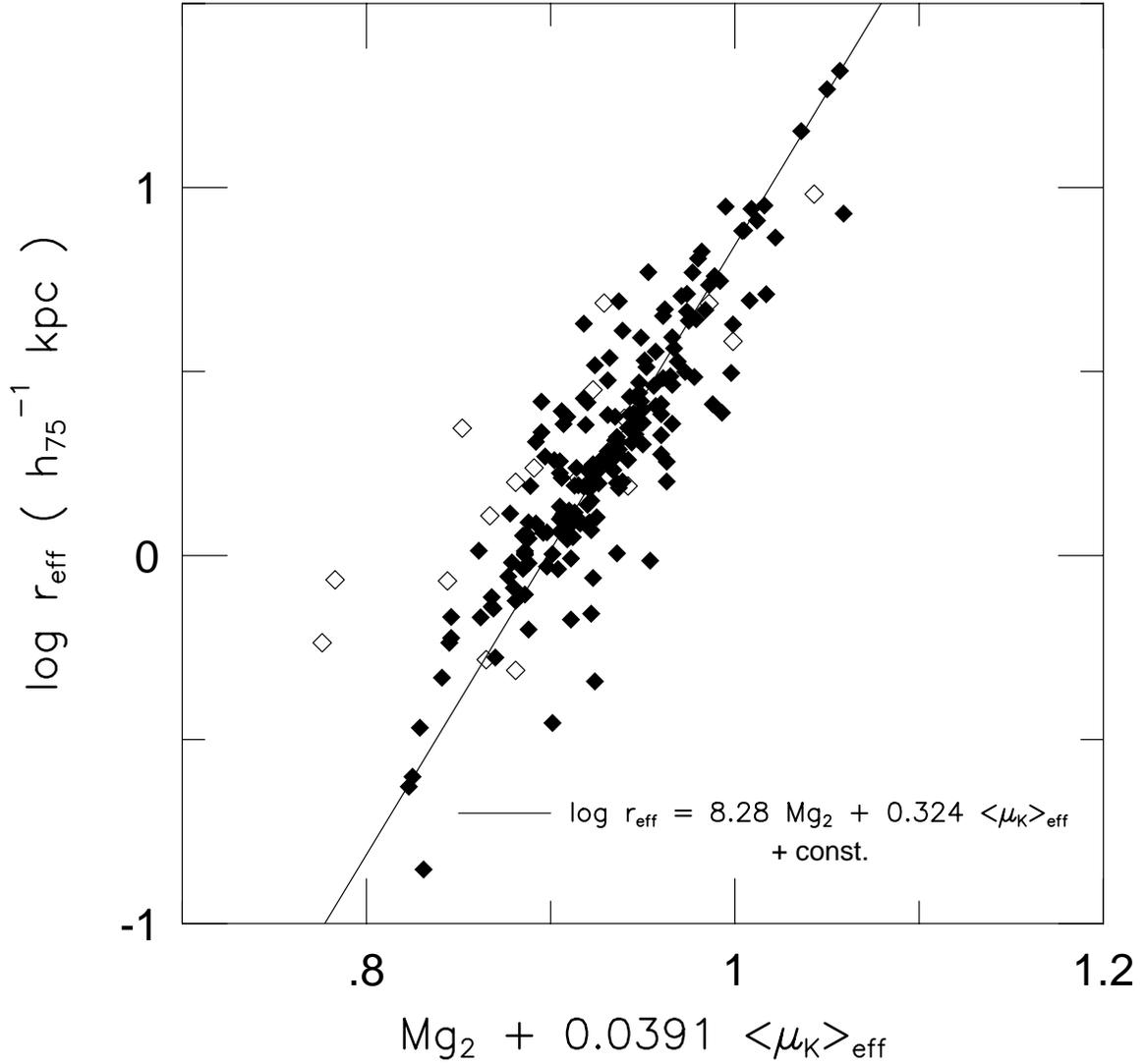}
	\caption[The \mgtwo\ Form of the Near--Infrared Fundamental Plane]{
	The near--infrared FP with the \mgtwo\ index substituted for the velocity dispersion.
	This figure is plotted to the same scale as Figure~\ref{fig-kfp-combined}(a), 
	hence a direct comparison of these two figures demonstrates how the scatter of
	the FP relation has increased by a factor of two by the substitution of \mgtwo\ for $\sigma_0$.
	Only a small part of this increase in scatter can be attributed to the larger measurement
	uncertainties of \mgtwo\ compared to $\sigma_0$, hence the correlation plotted in this
	figure cannot be an edge--on view of the FP.
	\label{fig-kfp-kfpmg2}
	}
\end{figure}

\placefigure{fig-kfp-kfpmg2}

If part of the scatter in the $\reff$--$\meanmueffK$--$\sigma_0$ FP can be attributed to differences
in stellar populations among elliptical galaxies, then the introduction of a stellar populations
``correction'' factor based on the \mgtwo\ index should be able to reduce the scatter of the FP even
though a small amount of additional observational uncertainty is added in the process.
The idea for this comes from the attempt by Guzm\'an \& Lucey (1993) to construct an ``age--independent''
distance indicator.
Here the method will be applied to the near--infrared data.

Using Bruzual (1983) evolutionary spectral synthesis models, Guzm\'an \& Lucey showed that
the effects of a burst of star formation involving 10\% of a galaxy's mass would appear as a
change of $\Delta m / \Delta$~\mgtwo\ roughly constant for times $\gtrsim 1$~Gyr or so after
the burst.
Hence, an offset in magnitude $\Delta m$ could be applied to each galaxy independently based
on its departure $\Delta$~\mgtwo\ from the \mgtwosigma\ relation.
Guzm\'an \& Lucey also showed that while the effect in the optical $V$--band was
$\Delta m_V / \Delta {\rm Mg}_2 \sim 10$, the effect in the near--infrared was much smaller at
\begin{equation}
	{ \Delta m_K \over \Delta {\rm~Mg}_2 } \sim 2 .
	\label{kfp-eq-kmag-over-mg2}
\end{equation}
This effect can basically be understood as a filling--in of the \mgtwo\ feature by the addition
of a continuum flux from hot, young stars.
The changes in the \mgtwo\ index are expected to be small, as \mgtwo\ is far more
sensitive to metallicity than it is to age or IMF (\cite{mould78}).

\begin{figure}
	\epsscale{1.0}
	\plotone{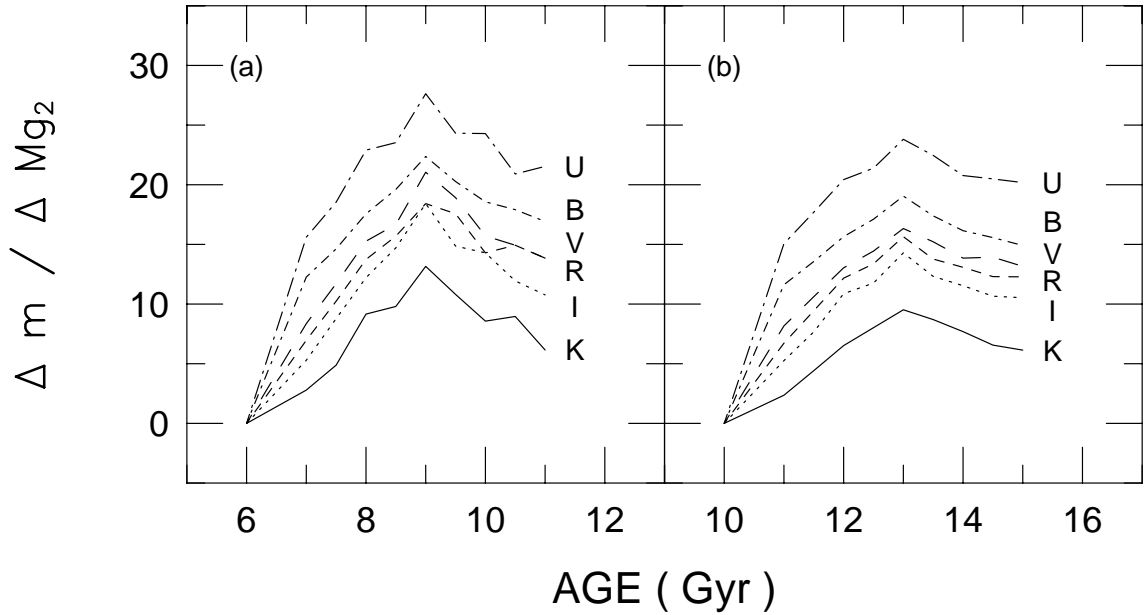}
	\caption[The Effects of a 10\% Young, 5~Gyr Population on Top of a 90\% Old Stellar Population]{
	The ratio of the change in magnitude $\Delta m$ to the change in the \mgtwo\
	index for the $UBVRIK$ bandpasses using the models of Worthey (1994).
	The two cases considered both have a 90\% (by mass) old stellar population
	component of 11~Gyr (a) and 15~Gyr (b) and [Fe/H]$=0$ in addition to a 10\%
	young stellar population that is 5~Gyr old at the present day.
	\label{fig-kfp-mg2-ubvrik}
	}
\end{figure}

The Guzm\'an \& Lucey procedure is repeated here for the Worthey (1994) stellar populations 
models in order to determine if the size of the predicted age effects are similar despite significant 
differences between the Bruzual (1983) and Worthey (1994) models.
The model adopted is similar to Guzm\'an \& Lucey in that it involves $10^6$~M$_\odot$ total mass,
90\% of which is 15~Gyr old in the present day, 10\% of which is 5~Gyr old, and all of which has solar metallicity.
A similar model is investigated that involves 90\% of the galaxy being 11~Gyr old in the present 
day and 10\% being 5~Gyr old.
The results from the Worthey models are plotted in Figure~\ref{fig-kfp-mg2-ubvrik} for the $UBVRIK$ bandpasses.


\placefigure{fig-kfp-mg2-ubvrik}

The Worthey models show modest agreement with
the Guzm\'an \& Lucey calculations based on Bruzual (1983) models for the optical bandpasses, although
the value of $\Delta$mag$ / \Delta$~\mgtwo\ is systematically 30--50\% higher.
In the $K$--band, however, Worthey's models are 3--4 times higher in this quantity.
The most likely explanation for this is the difficulty that many models have in producing enough
Mg relative to Fe for very metal--rich populations (see \cite{worthey92}, for example), although more fundamental
problems in the treatment of cool stellar atmospheres in the infrared for the Worthey models
(see \cite{chwobr96}) could also be relevant.
As a result of this discrepancy, the effects of a late burst of star formation involving a small
fraction of the galaxy mass on the FP cannot be assumed {\sl a priori}.

A better approach is to measure directly the possible contribution of younger stellar populations
using the FP itself, and then using the observations to constrain the models.
Combining Equations~\ref{kfp-eq-fpgen} and \ref{kfp-eq-mg2sigma}, defining 
$r_K = { \Delta m_K \over \Delta {\rm~Mg}_2 }$, and assuming that the mean surface brightness term 
in the FP is the only one affected by $r_K$ yields:
\begin{equation}
	\begin{array}{rcl}
	\log \reff ( {\rm arcsec} ) 	& = 	& a' \log\sigma_0 + b' \left( \meanmueffK - r_K \left[ {\rm~Mg}_2 
							- 0.173 \log\sigma_0 + 0.106 \right] \right) + c_i
	\end{array}
	\label{kfp-eq-kfp-mg2corr}
\end{equation}
where the primed coefficients are the ``age--independent'' form of the FP.
The Guzm\'an \& Lucey analysis predicts $r_K \sim 2$, while the Worthey models predict $r_K \sim 7$.

\begin{figure}
	\epsscale{1.0}
	\plotone{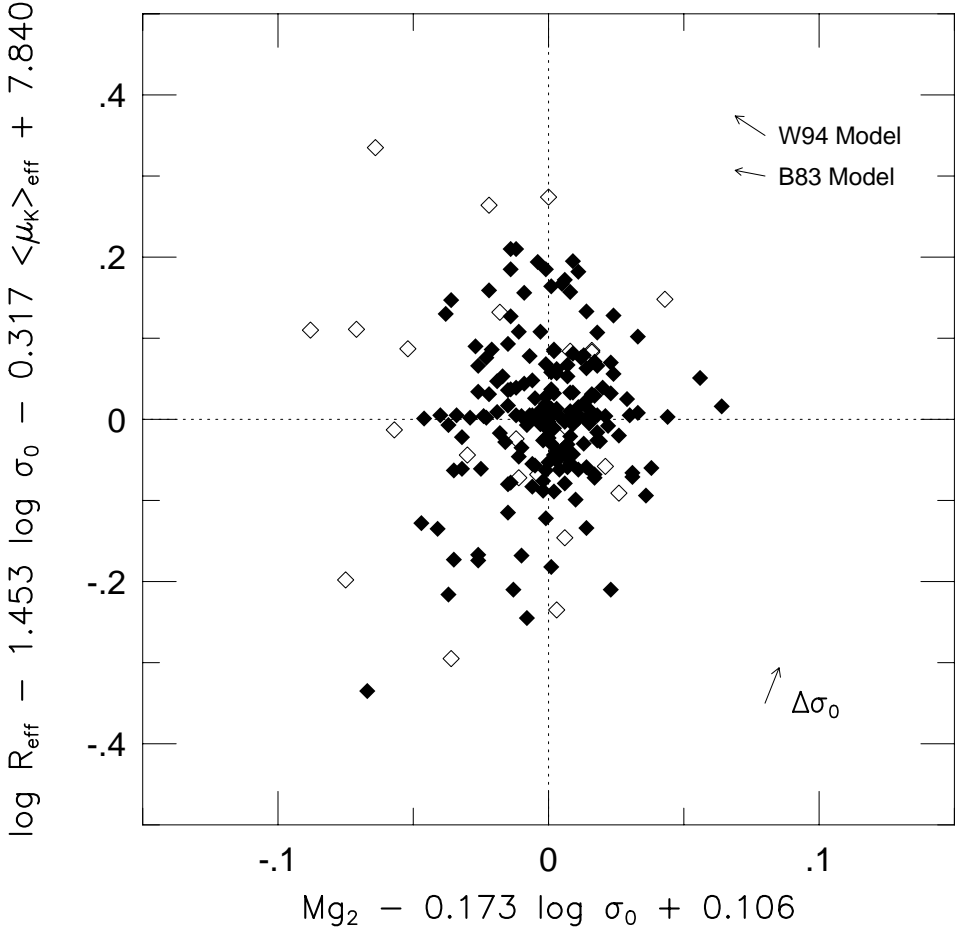}
	\caption[The Residuals of the Near--Infrared FP Plotted Against the Residuals of the
		\mgtwosigma\ Relation]{
	The residuals of the near--infrared FP plotted against the residuals of the \mgtwosigma\
	relation.
	For the purpose of this comparison, only those galaxies with good \mgtwo\ measurements 
	(see \S\ref{kfp-mg2-sigma}) were used to fit the near--infrared FP; the FP fit for these 182 galaxies is
	$\reff \propto \sigma_0^{1.45} \langle\Sigma_K\rangle_{\rm eff}^{-0.79}$.
	A correlation due to the measurement errors of $\log\sigma_0$ would act in the direction of the
	vectors in the lower--right of the panel labeled $\Delta\sigma_0$.
	If a galaxy had a starburst involving 10\% of its mass at 5~Gyr before the present day, this
	would produce an offset as shown for the Bruzual (1983; B83) and Worthey (1994; W94) models.
	Since there is no such correlation along these model vectors between the two sets of residuals, 
	this implies that there is no age effect, as traced by the \mgtwosigma\ relation, 
	which is the sole cause of the intrinsic scatter of the near--infrared FP.
	\label{fig-kfp-kfpmg2sigma-residuals}
	}
\end{figure}

Unfortunately, a minimization of Equation~\ref{kfp-eq-kfp-mg2corr} does not reveal an optimal value of $r_K$ since
values of $| r_K | >0$ increase the scatter of the equation due to the added measurement 
uncertainties.
Put another way, there was no significant improvement in the scatter (by $> 2$\%) of the FP for any 
$r_K$ in the range $-10 < r_K < +10$. 
Since the intrinsic thickness of the near--infrared FP is clearly resolved without including a 
\mgtwo\ term (i.e., Equation~\ref{kfp-eq-kfp-all} and Figure~\ref{fig-kfp-combined}), then some of that 
thickness could be due to variations in age among the stellar populations of elliptical galaxies.
This effect can be viewed from a similar perspective by attempting to correlate the residuals of the
near--infrared FP with the residuals of the \mgtwosigma\ relation.
This $\Delta$--$\Delta$ diagram is shown in Figure~\ref{fig-kfp-kfpmg2sigma-residuals}.
This figure shows no correlation among the residuals in the direction of the age vectors, implying that 
the resolved intrinsic scatter of both the near--infrared FP and the \mgtwosigma\ relations cannot
be caused by age effects alone.

\placefigure{fig-kfp-kfpmg2sigma-residuals}

Instead, insight can be gained by looking for changes in the relative distance modulus between a given
cluster and the Coma cluster.
When $r_K=2$, all 16 clusters and groups show changes in their distance moduli relative to Coma of
$\leq 0.05$~mag, which is smaller than the typical uncertainty of $0.1$~mag in the distance to a 
cluster of $N=20$ galaxies for a scatter of $\log\reff = 0.085$~dex.
The difference in relative distance modulus between the $r_K=0$ and $r_K=7$ cases, however, reach
as high as 0.12--0.18~mag in the case of several clusters (Virgo, Cen30, and Pegasus), which is
marginally significant.
Since the Worthey (1994) models probably overestimate $r_K$ due to their difficulty in producing
Mg for metal--rich populations, the conclusion is that $0.18$~mag is a firm upper limit to the effects
of age differences on the distance moduli derived using the near--infrared FP.
This result for the $K$--band is similar to that found in the $r$--band by J\o rgensen \etal\ (1996),
who found a minimally--significant contribution of $r_r = 1.3 \pm 0.8$.
They probably found no good correlation for the reason given above---that the increase of measurement
uncertainties as $r$ increases prevents effective minimization during the fitting---hence this may
not be a significant constraint on a superimposed intermediate age contribution.
Repeating the analysis using $r_V = 10$ for the Abell~2199 and Abell~2634 clusters for the galaxies 
in the $V$ and $K$ matched catalog 
produces a similarly small change in their distance moduli relative to Coma.
In summary, there is little evidence that adding a \mgtwo\ term (based on the \mgtwosigma\ relation) 
to the near--infrared FP to account for age differences in the ellipticals in different clusters 
causes a significant improvement over the zero--point for the relation and hence distances derived from it.

\begin{figure}
	\epsscale{1.0}
	\plotone{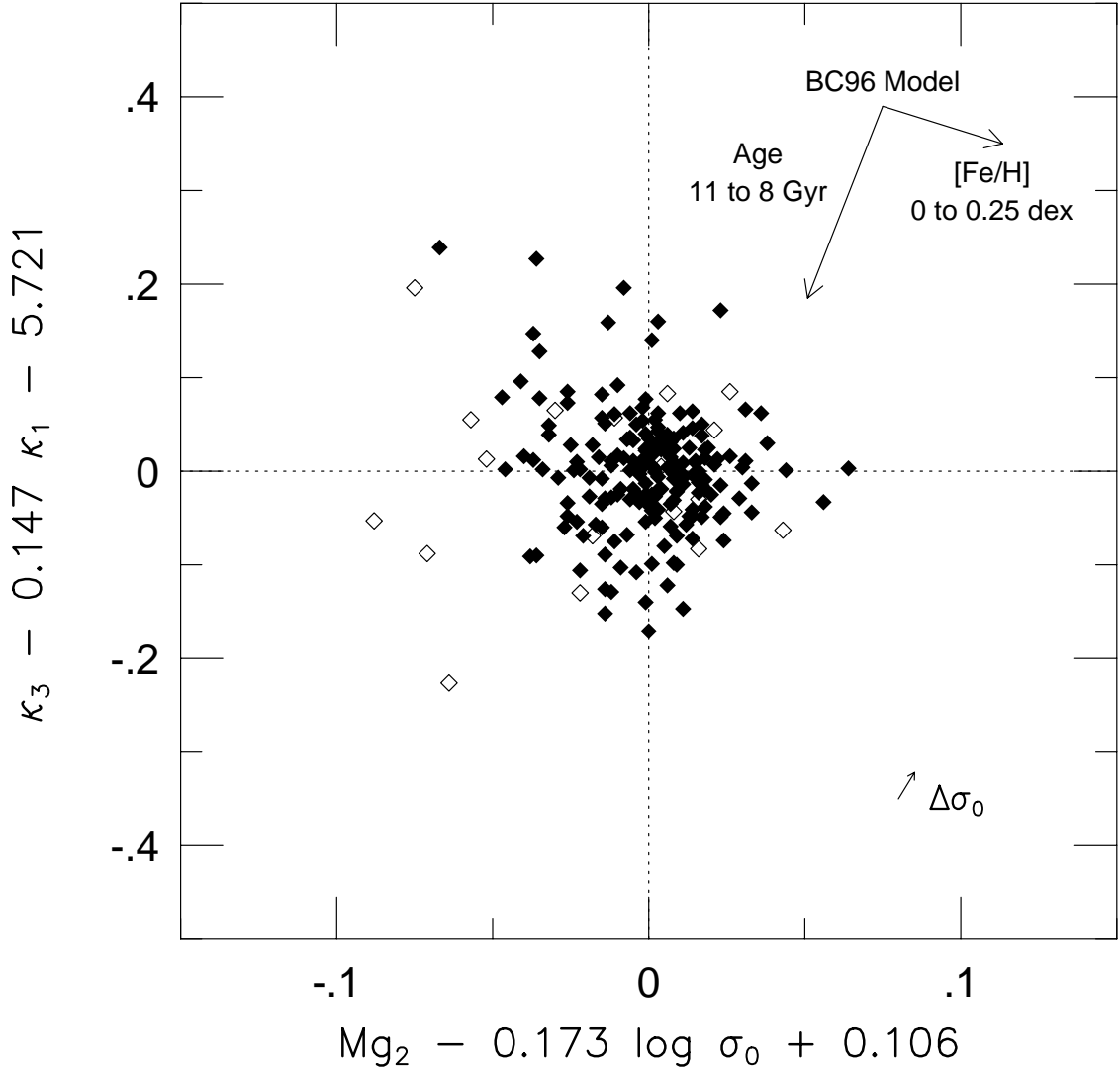}
	\caption[The Residuals from the \mgtwosigma\ and $\kappa_3$--$\kappa_1$ Relations]{
	The residuals from the \mgtwosigma\ relation (Equation~\ref{kfp-eq-kfp-mg2corr}) plotted 
	against the residuals from the $\kappa_3$--$\kappa_1$ relation (Equation~\ref{kfp-eq-kband-kappa}).
	Both relations show significant intrinsic scatter which could be due to variations in stellar populations
	at any given point on either relation.
	As shown by the vectors in the upper part of the figure, the Bruzual \& Charlot (1996) models show that
	this diagram is a powerful diagnostic for separating the effects of age and metallicity.
	These models show the effect for changing the age from 11 to 8~Gyr for a solar metallicity population,
	and for changing [Fe/H] from 0 to 0.25~dex in a 11~Gyr old population.
	The effect of correlated errors in $\sigma_0$ are shown in the lower--right of the figure.
	Variations in both age and metallicity at any given mass or luminosity appear to be necessary to 
	explain the intrinsic scatter in this diagram and the lack of correlation along either
	vector in the upper right of the figure.
	\label{fig-kfp-deltamg2sigma-deltakappa}
	}
\end{figure}

A similar approach could be pursued by looking at the correlation between the residuals of the
\mgtwosigma\ relation and the residuals of the $\kappa_3$--$\kappa_1$ relation.
If differences of \mgtwo\ at a fixed $\sigma_0$ indicate differences of stellar populations,
and if differences of $\kappa_3$ at a fixed $\kappa_1$ indicate differences in $(M/L)$ due to
stellar populations effects, then the residuals of these two relations
(Equations~\ref{kfp-eq-kfp-mg2corr} and \ref{kfp-eq-kband-kappa}) should correlate in
a manner that is consistent with stellar populations effects.
These residuals are plotted against each other in Figure~\ref{fig-kfp-deltamg2sigma-deltakappa}.

\placefigure{fig-kfp-deltamg2sigma-deltakappa}

This figure is an excellent diagnostic for distinguishing between age and metallicity effects,
since \mgtwo\ decreases for younger stellar populations while it increases for higher metallicities,
but $(M/L_K)$ decreases for both.
[If the vertical axis were $\Delta (M/L)$ measured at any optical wavelength, then age and metallicity would
instead act nearly parallel.]
The lack of any preferred correlation along either the age or metallicity vectors in 
Figure~\ref{fig-kfp-deltamg2sigma-deltakappa}, while at the same time having a substantial 
intrinsic scatter for both relations, strongly indicates that both age and metallicity 
variations exist at any given point on both the \mgtwo\ and $\kappa_3$--$\kappa_1$ relations.

The H$\beta$ index is expected to be a good indicator of the presence of a young stellar component,
and either the \mgtwo\ or $\log\langle$Fe$\rangle$ indices should be good indicators of the
mean metallicity of the stellar content.
The large intrinsic scatter between the H$\beta$ and \mgtwo\ indices, 
the lack of a correlation altogether between H$\beta$ and $\log\langle$Fe$\rangle$, 
the strong correlation between \mgtwo\ and $\sigma_0$, 
and the weak correlation between $\log\langle$Fe$\rangle$ and $\sigma_0$ (\cite{jorgensen98})
all indicate that there exist significant variations in both age and metallicity
for any given value of $\sigma_0$.
This is fully consistent with the argument above based on the residuals of the \mgtwosigma\ and
$\kappa_3$--$\kappa_1$ relations.

\subsection{The Faber--Jackson Relation \label{kfp-fj}}

\begin{figure}
	\epsscale{0.9}
	\plotone{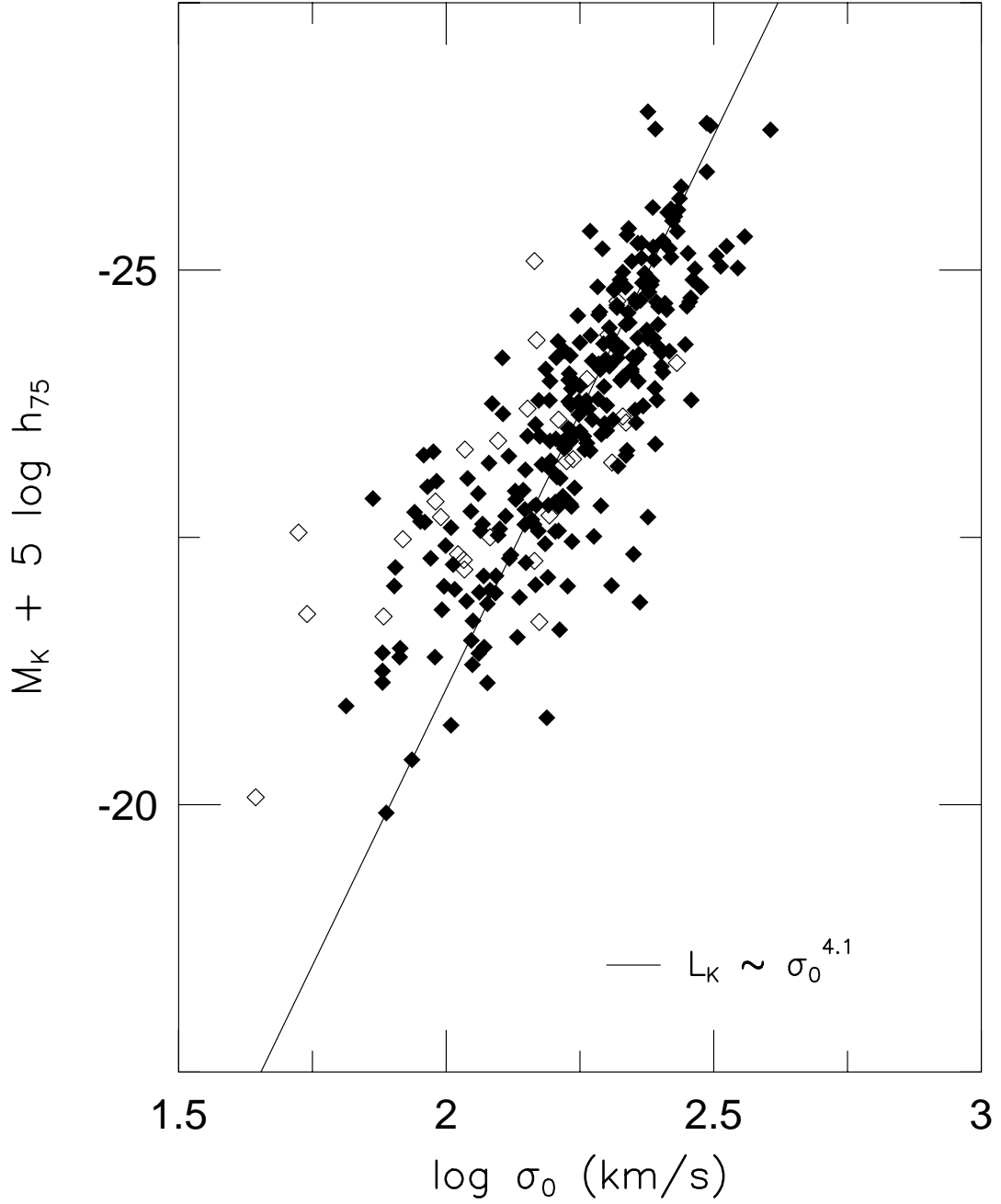}
	\caption[The Faber--Jackson Relation]{
	The Faber--Jackson relation between luminosity and central velocity dispersion.
	The best--fitting relation is $L_K \propto \sigma_0^{4.14 \pm 0.22}$ with a large
	scatter of $0.93$~mag.
	The scatter is significantly smaller in the Coma cluster at $0.72$~mag.
	\label{fig-kfp-fj}
	}
\end{figure}

The correlation between luminosity and central velocity dispersion for elliptical galaxies
was first noticed by Faber \& Jackson (1976).
If we fit the relation $L \propto \sigma_0^a$, this is equivalent to fitting
$\Ktot = - 2.5 a \log \sigma_0 + b$.
The best fitting Faber--Jackson relation for the $K$--band data is
\begin{equation}
	\begin{array}{rclllr}
	M_K = \Ktot - 34.91 + 5 \log h_{75} & = &\, \, -10.35 \log \sigma_0 & & N = 252 & {\rm rms} = 0.93 {\rm~mag}  \\
					    &   &    \pm 0.55
	\end{array}
	\label{kfp-eq-fj}
\end{equation}
assuming $H_0 = 75$~km~s$^{-1}$ and that the Coma cluster ($cz=7200$~km~s$^{-1}$) is at rest with
respect to the Hubble flow.
The relation is plotted in Figure~\ref{fig-kfp-fj}.
The scatter of this relation is significantly smaller in the Coma cluster alone (rms$= 0.72$~mag).

\placefigure{fig-kfp-fj}

\subsection{The Modified Faber--Jackson Form of the FP \label{kfp-modified-fj}}

Since there is substantial scatter in the Faber--Jackson relation due to variations
in surface brightness among galaxies at a given luminosity and central velocity
dispersion, an alternate form of the FP is to substitute $\Ktot$ for $\reff$.
This will be referred to as the ``modified Faber--Jackson'' relation, as it adds the
additional $\meanmueff$ term to the Faber--Jackson relation.
The form of this equation is
\begin{equation}
	\Ktot = a' \log \sigma_0 ({\rm km~s}^{-1}) + b' \meanmueff ({\rm mag~arcsec}^{-2}) + c'_i
	\label{kfp-eq-fpgen-modified-fj}
\end{equation}
where the primed coefficients are used here for the modified Faber--Jackson relation.
Since in the case of a pure de Vaucouleurs profile $\Ktot({\rm mag}) = -5 \log \reff + \meanmueffK - 1.995$, 
the value $\log\reff = a \log\sigma_0 + b \meanmueffK + {\rm~constant}$ can be substituted for
$\reff$ resulting in $a = -5 a'$ and $b = 0.2 ( 1 - b' )$, thereby relating 
Equations~\ref{kfp-eq-fpgen} and \ref{kfp-eq-fpgen-modified-fj} to each other.
The best--fitting relation of the modified Faber--Jackson form of the FP is:
\begin{equation}
	\begin{array}{rcllll}
	M_K + 5 \log h_{75}  & = &\, \, -8.16 \log \sigma_0 - &\, \, 0.585 \meanmueffK + c_i & N=251 & {\rm rms} = 0.51 {\rm~mag} \\
	                          &   &  \pm 0.47                 &  \pm 0.062
	\end{array}
	\label{kfp-eq-modified-fj}
\end{equation}
which has a scatter only 10\% larger than the standard form of the near--infrared FP given in
Equation~\ref{kfp-eq-kfp-all}.
This equation represents the scaling relation
$L_K \propto \sigma_0^{3.26 \pm 0.19} \langle\Sigma_K\rangle_{\rm eff}^{-0.59 \pm 0.06}$,
which is fully equivalent within the uncertainties to the standard form of the near--infrared FP.
Since the uncertainty on $\Ktot + 0.60 \meanmueffK$ is $0.068$~mag, the total observational 
uncertainties in Equation~\ref{kfp-eq-kfp-all} and \ref{kfp-eq-modified-fj} 
are similar.
Hence Equation~\ref{kfp-eq-modified-fj} also shows substantial intrinsic scatter in the properties 
of elliptical galaxies at any point along the FP.

\begin{figure}
	\epsscale{1.0}
	\plotone{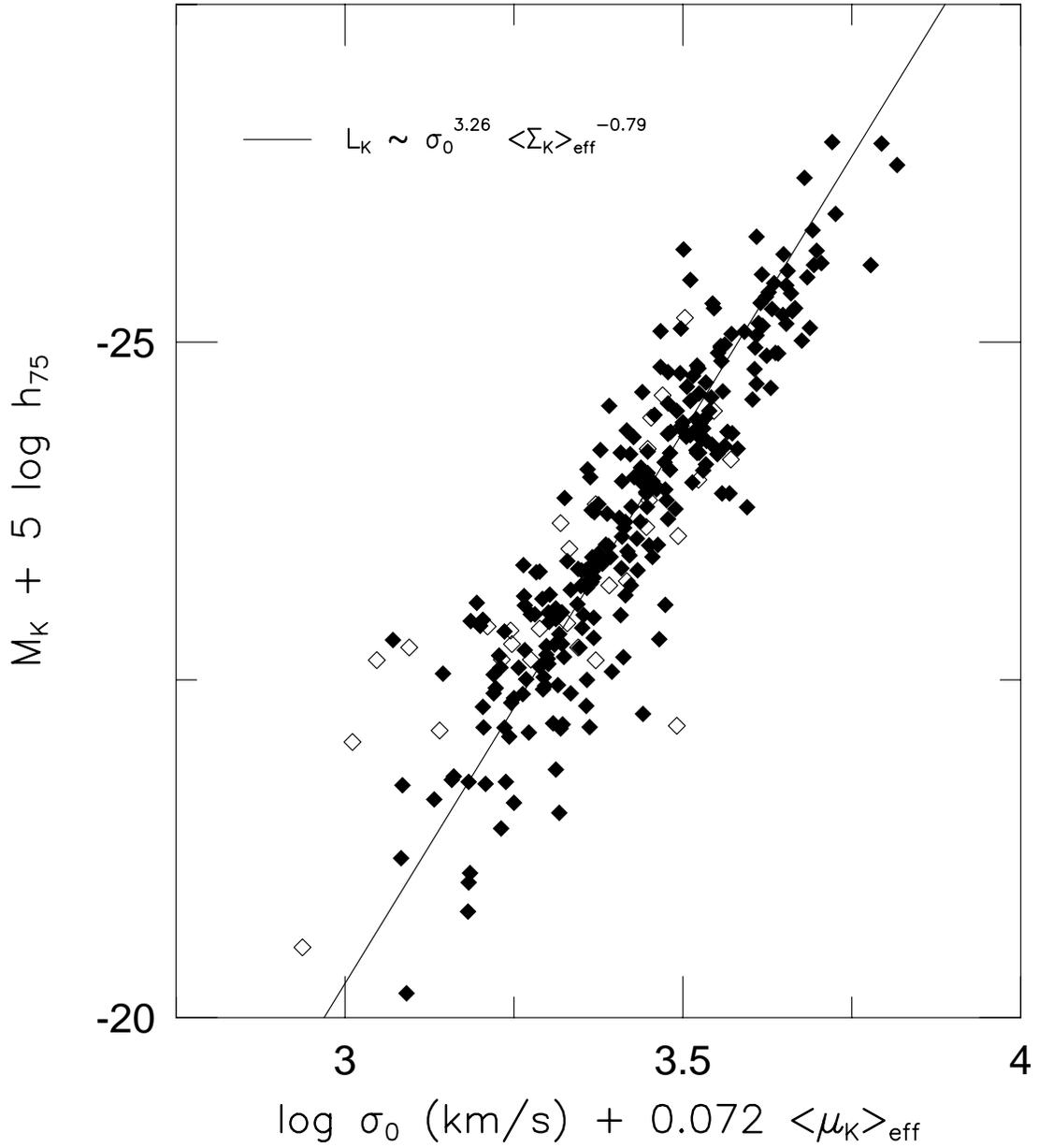}
	\caption[The Modified Faber--Jackson Relation Form of the FP]{
	The ``modified Faber--Jackson'' form of the FP.
	The best--fitting relation is 
	$L_K \propto \sigma_0^{3.26 \pm 0.19} \langle\Sigma_K\rangle_{\rm eff}^{-0.59 \pm 0.06}$
	with a scatter of $0.51$~mag.
	This form of the FP is nearly identical to that of Figure~\ref{fig-kfp-combined} 
	and Equation~\ref{kfp-eq-kfp-all}, although it shows 10\% larger scatter primarily due
	to larger observational uncertainties.
	\label{fig-kfp-modified-fj}
	}
\end{figure}

\placefigure{fig-kfp-modified-fj}

\subsection{The Kormendy Relation \label{kfp-kormendy}}

\begin{figure}
	\epsscale{1.0}
	\plotone{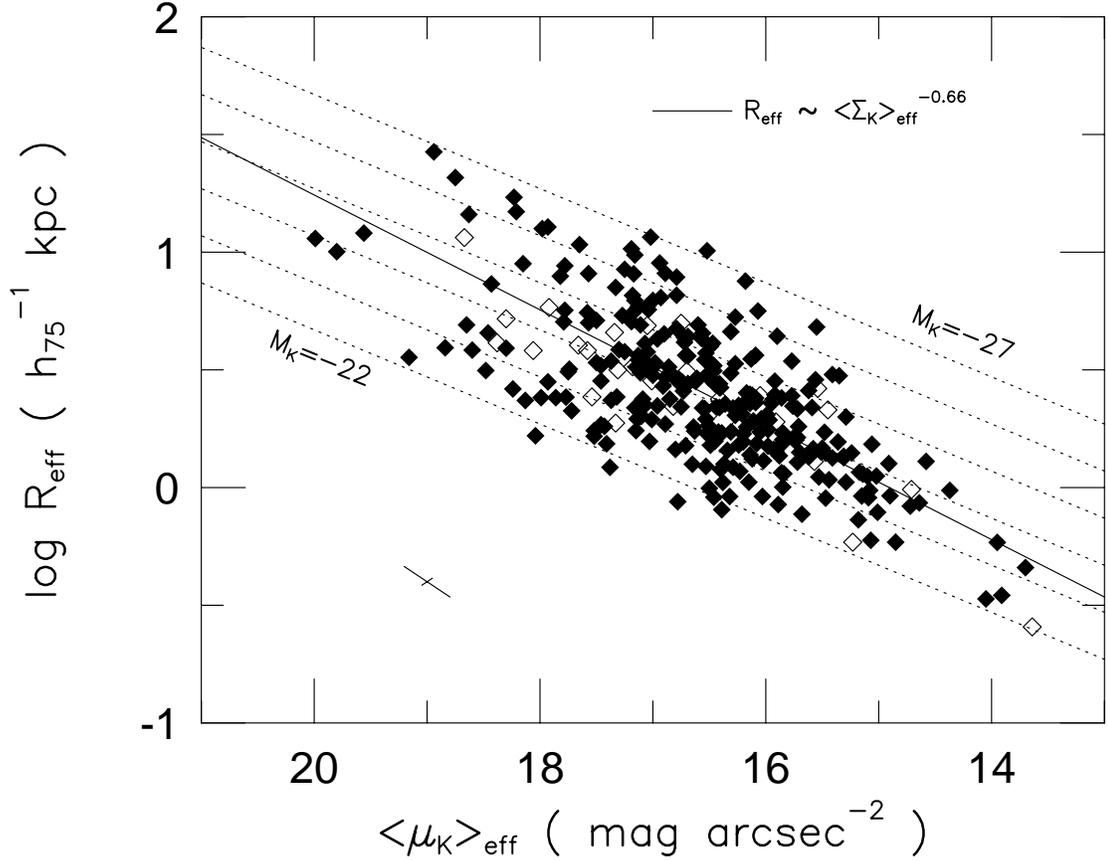}
	\caption[The Kormendy Relation Between Effective Radius and Mean Surface Brightness]{
	The Kormendy relation between effective radius and mean surface brightness.
	The best--fitting relation is $\Reff \propto \langle\mu_k\rangle_{\rm eff}^{-0.61 \pm 0.07}$ with a
	scatter of $0.227$~dex in $\log\Reff$, which is 2.4 times worse than the scatter of the FP
	(which has the additional $\sigma_0$ term).
	The scatter is significantly smaller in the Coma cluster at $0.198$~dex.
	Measurement errors in $\Reff$ and $\meanmueff$ are correlated nearly along the relation, as shown
	by the representative error bars in the lower--left of the figure. 
	The measurement uncertainties
	perpendicular to the relation are only $0.015$~dex in $\log\Reff - 0.32\meanmueffK$.
	Luminosity changes act perpendicular to the relation, and representative lines of $M_K = -22$~mag
	to $M_K = -27$~mag are shown as dotted lines.
	\label{fig-kfp-kormendy}
	}
\end{figure}

Effective radius and mean surface brightness, which are two of the three terms in the FP,
are correlated with each other (\cite{kormendy77}).
The best fitting Kormendy relation for the $K$--band data is
\begin{equation}
	\begin{array}{rclll}
	\log \Reff h_{75}^{-1}       & = &\, \, 0.244 \meanmueffK - 3.637 & N = 269 & {\rm rms} = 0.227 {\rm~dex}  \\
				     &   &  \pm 0.029
	\end{array}
	\label{kfp-eq-kormendy}
\end{equation}
assuming $H_0 = 75$~km~s$^{-1}$ and that the Coma cluster ($cz=7200$~km~s$^{-1}$) is at rest with
respect to the Hubble flow.
The relation is plotted in Figure~\ref{fig-kfp-kormendy}.
The scatter of this relation is significantly smaller in the Coma cluster alone (rms$= 0.198$~dex).
Measurement errors in $\reff$ and $\meanmueff$ are correlated and act in a direction nearly
parallel to the Kormendy relation, but are not nearly large enough to account for the spread in
galaxy properties along the relation.
Changes in luminosity are skewed with respect to this relation and shown in Figure~\ref{fig-kfp-kormendy}.
For this reason it is necessary that a magnitude--limited sample be defined in a consistent
way for all clusters studied before conclusions based on changes in the zero--point (due to distances
for nearby clusters or evolutionary brightening for higher redshifts) can be made.

\placefigure{fig-kfp-kormendy}

\subsection{The Radius--Luminosity Relation \label{kfp-radius-lum}}

\begin{figure}
	\epsscale{1.0}
	\plotone{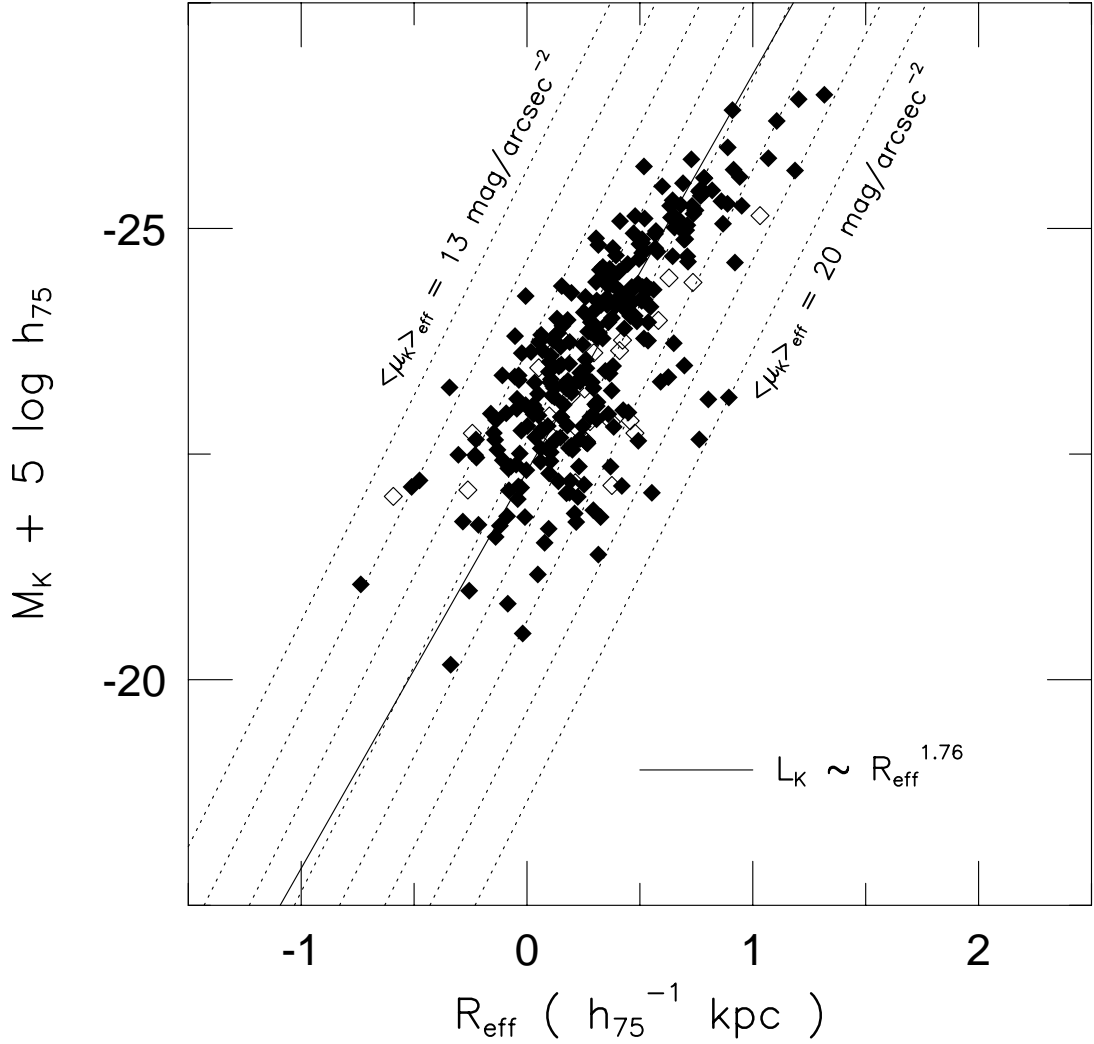}
	\caption[The Relation Between Luminosity and Effective Radius]{
	The relation between $K$--band total luminosity and effective radius given by
	$L_K \propto \Reff^{1.76 \pm 0.10}$.
	There is substantial intrinsic scatter in this relation due to variations in surface brightness.
	The dotted lines each represent the variation of $M_K$ and $\Reff$ at constant surface brightness.
	\label{fig-kfp-magradius}
	}
\end{figure}

The correlation between the effective radius and total magnitude for elliptical galaxies
has long been used for distance scale work and especially cosmological tests
(see \cite{sandageperl2} and references therein).
The best fitting radius--luminosity relation for the $K$--band data is
\begin{equation}
	\begin{array}{rclll}
	M_K + 5 \log h_{75} & = &\, \, -4.40 \Reff - 22.31 & N = 269 & {\rm rms} = 0.88 {\rm~mag}  \\
			    &   &   \pm 0.26
	\end{array}
	\label{kfp-eq-magradius}
\end{equation}
assuming $H_0 = 75$~km~s$^{-1}$ and that the Coma cluster ($cz=7200$~km~s$^{-1}$) is at rest with
respect to the Hubble flow.
The relation is plotted as Figure~\ref{fig-kfp-magradius}.
There is intrinsic scatter to this relation that is a result of the variation in surface brightness
at a given radius and luminosity; lines of constant surface brightness are plotted in the figure
to demonstrate this effect.

\placefigure{fig-kfp-magradius}

\section{Exploring Simple Models for the Origins of the Elliptical Galaxy Scaling Relations 
	in the Near--Infrared
	\label{kfp-simple} }

The near--infrared FP has been shown in \S\ref{kfp-kfp} to be represented by the scaling relation
$\reff \propto \sigma_0^{1.53 \pm 0.08} \langle\Sigma_K\rangle_{\rm eff}^{-0.79 \pm 0.03}$.
This relation shows a significant deviation from the optical forms of the FP:
$\reff \propto \sigma_0^{1.24 \pm 0.07} \langle\Sigma\rangle_{\rm eff}^{-0.82 \pm 0.02}$
(\cite{jfk96}) and
$\reff \propto \sigma_0^{1.38 \pm 0.04} \langle\Sigma\rangle_{\rm eff}^{-0.82 \pm 0.03}$
(\cite{hudson97}) in the $R$--band; or
$\reff \propto \sigma_0^{1.13} \langle\Sigma\rangle_{\rm eff}^{-0.79}$ in the $V$--band
(\cite{guzman93b}).
There are two simple conclusions to draw from these data:  (1) the slope of the near--infrared
FP deviates from the virial expectation of $\reff \propto \sigma_0^{2} \langle\Sigma\rangle_{\rm eff}^{-1}$,
and (2) the slope of the FP increases with wavelength.
A third insight derives from the fact that the scatter of the FP is very similar at
all wavelengths.
These three points are sufficient to discuss several simple models for the physical
origins of the FP.

The age--metallicity model of Worthey, Trager, \& Faber (1995)---based on the form of the FP
in the optical, various line indices, and simple stellar populations model comparisons---incorrectly 
predicts that the near--infrared FP should follow the virial form.
Another model, that the FP slope is caused by deviations of the velocity distributions of elliptical 
galaxies from a homologous scaling family (Capelato \etal\ 1995), cannot account for the variations of
the slope of the FP with wavelength.
If this breaking of homology has its origin in dissipation-less merging, then this effect also cannot explain the
correlation between \mgtwo\ and $\sigma_0$.
A final model, which suggests that deviations of the light distributions of elliptical galaxies from the 
de Vaucouleurs $r^{1/4}$ form is the cause of the FP slope, is unable to account for the slope of the 
FP in the optical (\cite{graham97}), and for the same reasons it cannot explain the near--infrared 
FP slope.

The deviation of the slope of the near--infrared FP from the virial expectation, assuming homology and
constant $M/L$ among ellipticals, is a very significant result.
This requires a breakdown of one or both assumptions:  either $M/L$ is systematically varying along
the FP, or elliptical galaxies are systematically deviating from a homologous scaling family.
If age is the stellar populations parameter which causes variations in the slope of the FP with wavelength, 
then age alone might possibly produce the slope of the $K$--band FP.
This conclusion, however, is severely limited by the possibility of homology breaking along the elliptical
galaxy sequence.

Allowing for structural deviations from homology, in the form of a Sersic $r^{1/n}$ profile, does not appear to
cause significant changes to the slope of the FP for high S/N, $V$--band data in the Virgo cluster (\cite{graham97}).
Instead, allowing for dynamical deviations from homology, via galaxy to galaxy variations in the mapping from 
$\sigma_0$ to $\sigma_{\rm eff}$, appears to cause significant changes in the slope of the FP (\cite{busarello97};
cf. \cite{graham97}).
Busarello \etal\ found a relationship between the velocity dispersions to be
$\log\sigma_0 = ( 1.28 \pm 0.11 ) \log\sigma_{\rm eff} - 0.58$.
Substituting for $\log\sigma_0$ into the $K$--band FP solution in Equation~\ref{kfp-eq-kfp-all} produces
\begin{equation}
\begin{array}{rcll}
\log \reff & = &\, \, 1.96 \log \sigma_{\rm eff} + &\, \, 0.314 \meanmueffK + {\rm~constant} \\
           &   &  \pm 0.20                 	   &  \pm 0.011
\end{array}
\label{kfp-eq-dyn-nonhom}
\end{equation}
which is statistically consistent with the virial expectation of $\reff \propto \sigma^2$.
This argument suggests that the deviation of the near--infrared FP from the virial expectation
can be fully explained by systematic deviations of the velocity structure of elliptical
galaxies from a homologous family, removing the requirement of large age spreads
among elliptical galaxies.

Since either dynamical non-homology or large age spreads could produce the slope of the
near--infrared FP, it is impossible to distinguish between these two simple models without
further analysis.
In addition, any model which incorporates either age or dynamical deviations from a homology
along with metallicity variations cannot be excluded in this simple analysis, either.
This strongly suggests that a much more detailed analysis, along with a more complicated model with
several different variables, is necessary to explain the global properties of elliptical galaxies.

One last {\sl ad hoc} model can be constructed in which there is a \emph{conspiracy} between
metallicity and age effects that act in a manner to keep the FP thin.
In this model, there can be a large spread in age and metallicity at any given point of the FP---under
the constraint that the two effects of age and metallicity work opposite to each other and thereby
cancel out to maintain a small scatter.
While this model would work at optical wavelengths, the independence of near--infrared light to
metallicity would cause the thinness of the optical FP to break down into a thick, 
near--infrared FP.
Since the near--infrared FP has similar observed and intrinsic thicknesses when compared to the optical
FP, especially when considering the additional observational uncertainties on $\reff$ caused by
$\sigma_0$ for the steeper slope of the near--infrared FP, this model can be excluded.

\section{Discussion \label{kfp-discussion} }

The near--infrared FP that has been constructed in this paper has several important properties:
(1) it deviates from the virial expectation (assuming constant $M/L$ and homology);
(2) it is steeper than the optical FP relations; (3) it has a similarly small scatter when compared
to the optical FP relations; and (4) it has a small, but significantly resolved, intrinsic scatter.
These observational constraints are sufficient to exclude a number of simple models for
the origin of the FP, but they do not provide unique discrimination between composite models
which include either age, systematic deviations from dynamical homology, or both.
Small additional contributions due to metallicity variations are also possible.

Better insight is gained by including the \mgtwo\ index into the analysis.
The \mgtwo\ form of the FP has much larger scatter than the standard form, which argues that 
\mgtwo\ does not uniquely specify the depth of the potential well for each galaxy.
This is entirely consistent with the resolved intrinsic scatter of the \mgtwosigma\ relation.
If some physical process like galactic winds (\cite{yoshii87}) caused the metallicity
and potential well for all elliptical galaxies to behave like a one--parameter family,
then some other physical property, such as dissipation-less mergers or a large scatter in
formation times, would be required to produce the small intrinsic scatter of the \mgtwosigma\ 
relation and the large intrinsic scatter of the \mgtwo\ form of the FP.

The near--infrared FP has the unique property that the $K$--band light is virtually independent
of metallicity.
For this reason, residuals of the $K$--band FP (or the $K$--band relationship between
$\kappa_1$ and $\kappa_3$) and the \mgtwosigma\ relations could provide a strong discrimination 
between age and metallicity effects.
The lack of any clear correlation between these residuals implies that neither age nor metallicity
is a unique contributor to the intrinsic scatter of the FP or the \mgtwosigma\ relations.

The Fundamental Plane is not just a simple correlation of the observed properties of elliptical galaxies,
but rather a unique tool for studying the intrinsic physical properties spanned by these galaxies.
The remarkable homogeneity of properties of elliptical galaxies that is implied by the regularity
and thinness of the optical FP is clearly reproduced by their similarly regular properties in
the near--infrared bandpass.

\acknowledgments

J. Blakeslee and J. Tonry are thanked for providing their SBF distance moduli in electronic form.
This research has made use of the NASA/IPAC Extragalactic Database (NED)
which is operated by the Jet Propulsion Laboratory, California Institute
of Technology, under contract with the National Aeronautics and Space
Administration.
During the course of this project, M.~A.~P. received financial support from Jesse
Greenstein and Kingsley Fellowships, and Hubble Fellowship grant HF-01099.01-97A from 
STScI (which is operated by AURA under NASA contract NAS5-26555); 
S.~G.~D. was supported in part by grants from the NSF (AST--9157412) and the
Bressler Foundation.


\clearpage


\makeatletter
\def\jnl@aj{AJ}
\ifx\revtex@jnl\jnl@aj\let\tablebreak=\nl\fi
\makeatother
\scriptsize
\begin{deluxetable}{lccccccccccccc}
\tablewidth{0pc}
\tablecaption{Fits for Each Cluster or Group for the Near--Infrared FP
	\label{tab1} }
\tablehead{
Cluster or	& \multicolumn{3}{c}{Simultaneous Fit}&	& \multicolumn{5}{c}{Individual Fits} 			
	& 	& \multicolumn{3}{c}{Constrained $b=0.314$ Fits}\\
\cline{2-4} \cline{6-10} \cline{12-14}
Group		& $c_i$		& $N$	& rms 	      & & $a$	& $\Delta a$	& $b$	& $\Delta b$	& rms	
	& 	& $a$	& $\Delta a$	& rms \\
		&		&	& (dex)       & &	&		&	&		& (dex) 
	&	&	&		& (dex) \\
(1)		& (2)		& (3)	& (4)	      & & (5)	& (6)		& (7)	& (8)	&	 (9)	
	&	& (10)	& (11)		& (12)
}

\startdata
Coma         	&    -7.950    	&  60  	& 0.086       & & 1.33  & 0.19		& 0.302	& 0.03		& 0.082 
	&	& 1.57	& 0.15		& 0.088 \\
A194         	&    -7.734    	&  16  	& 0.107       & & 1.57  & 0.21		& 0.254	& 0.05		& 0.106 
	&	& 1.60	& 0.16		& 0.110 \\
A2199        	&    -8.128    	&  23  	& 0.093       & & 1.53  & 0.22		& 0.342	& 0.03		& 0.088 
	&	& 1.40	& 0.16		& 0.086 \\
A2634        	&    -8.028    	&  15  	& 0.076       & & 1.19  & 0.74		& 0.292	& 0.06		& 0.061 
	&	& 1.24	& 0.30		& 0.063 \\
Cen45        	&    -7.543    	&   6  	& 0.071       & &\nodata & \nodata	&\nodata & \nodata	& \nodata 
	&	&\nodata & \nodata	& \nodata \\
Cen30        	&    -7.526    	&  14  	& 0.124       & & 1.72  & 0.41		& 0.299	& 0.08		& 0.123 
	&	& 2.05	& 0.39		& 0.139 \\
Fornax       	&    -7.274    	&  15  	& 0.137       & & 2.56  & 0.65		& 0.339	& 0.06		& 0.156 
	&	& 2.11	& 0.30		& 0.128 \\
Hydra        	&    -7.669    	&  17  	& 0.086       & & 1.76  & 0.34		& 0.344	& 0.03		& 0.080 
	&	& 1.77	& 0.15		& 0.086 \\
Klemola~44   	&    -8.041    	&  11  	& 0.067       & & 1.50  & 0.58		& 0.309	& 0.05		& 0.068 
	&	& 1.74	& 0.28		& 0.069 \\
Pegasus      	&    -7.580    	&   4  	& 0.048       & &\nodata & \nodata	&\nodata & \nodata	& \nodata 
	&	&\nodata & \nodata	& \nodata \\
Perseus      	&    -7.802    	&  19  	& 0.100       & & 1.98  & 0.59		& 0.310	& 0.05		& 0.125 
	&	& 1.66	& 0.33		& 0.104 \\
Pisces       	&    -7.723    	&  11  	& 0.087       & & 1.04  & 0.27		& 0.350	& 0.05		& 0.055 
	&	& 1.17	& 0.19		& 0.055 \\
Virgo        	&    -7.175    	&  27  	& 0.115       & & 1.77  & 0.25		& 0.374	& 0.03		& 0.120 
	&	& 1.62	& 0.12		& 0.118 \\
Eridanus     	&    -7.312    	&   5  	& 0.061       & &\nodata & \nodata	&\nodata & \nodata	& \nodata 
	&	&\nodata & \nodata	& \nodata \\
Leo          	&    -6.932    	&   3  	& 0.253       & &\nodata & \nodata	&\nodata & \nodata	& \nodata 
	&	&\nodata & \nodata	& \nodata \\
N5846grp     	&    -7.436    	&   5  	& 0.101       & &\nodata & \nodata	&\nodata & \nodata	& \nodata 
	&	&\nodata & \nodata	& \nodata \\
\tablecomments{
(1)  The FP fits in this table are to the form of Equation~\ref{kfp-eq-fpgen}.
(2)  The simultaneous fit for columns 2--4 corresponds to the solution in Equation~\ref{kfp-eq-kfp-all},
	allowing only the intercepts $c_i$ to vary between clusters.
(3)  The individual cluster FP fits in columns 5--9 are for only those 11 clusters with numbers of
	galaxies $N \geq 10$.
(4)  The constrained, individual cluster fits of columns 10--12 were obtained by fixing $b = 0.314$.
(5)  The rms in all cases is evaluated along the $\log\reff$ axis.
}
\enddata
\end{deluxetable}
\normalsize


\clearpage


\clearpage


\end{document}